\documentstyle[twocolumn,aps]{revtex}

\begin{document}
\draft
\title{Statistics of Rare Events in Disordered Conductors.}
\author{Igor E. Smolyarenko$^{a,b}$ and Boris L. Altshuler$^b$}
\address{$^a$Lyman Laboratory of Physics, Harvard University, Cambridge, MA
02138 \and \\ and \\$^b$NEC Research Institute, 4 Independence Way, Princeton,
NJ 08540} 
\date{\today}
\maketitle

\begin{abstract}
Asymptotic behavior of the distribution functions of local
quantities in disordered conductors is studied in the weak disorder limit by
means of an optimal fluctuation method. It is argued that this method is more
appropriate for the study of seldom occurring events than the approaches based
on  nonlinear $\sigma$-models because it is capable of correctly handling 
fluctuations of the random potential with large amplitude as well as the
short-scale structure of the corresponding solutions of the Schr\"{o}dinger
equation. For two- and three-dimensional conductors new asymptotics of the
distribution functions are obtained which in some cases differ significantly from
previously established results.
\end{abstract}

\section{Introduction.}

It has been well known for at least a decade \cite{redbook} that an adequate
description of the Anderson localization transition in disordered mesoscopic
conductors must necessarily be formulated in terms of the full distribution
functions of conductance $g$ and/or other quantities characterizing the sample.
As a consequence, even well into the metallic regime it should be possible  to
observe the onset of localization by studying deviations of asymptotic tails of
the distribution functions from their behavior in the infinite conductance
limit. Such a study was first performed in Ref. \cite{AKL} within the
framework of the replica-based nonlinear $\sigma$-model
\nolinebreak\cite{Wegner}.  It was
demonstrated that the tails of the distributions of such quantities as
conductance, local density of states, current relaxation times, etc. are all
described in two spatial dimensions by rather similar logarithmically normal
asymptotes whereas Gaussian distributions would be expected for
$g\rightarrow\infty$. Recently, it was discovered \cite{Kh1} that these
asymptotes can be obtained in a more straightforward and elegant manner using
the optimal fluctuation method in conjunction with the supersymmetric variant
of the nonlinear $\sigma$-model \cite{Efetov}. In a series of subsequent
publications the method was employed to study statistics of eigenfunction
amplitudes in weakly localized two-dimensional conductors \cite{EfFal1} and
then extended to systems in three spatial dimensions \cite{EfFal2,Mirlin,Kh3}.

The common feature of all the results obtained so far with the use of various
$\sigma$-models is that asymptotic behavior of a distribution function ${\cal
P}\left(t\right)$ for large $t$  has the following form for two- and 
three-dimensional systems:
\begin{equation}
\label{eq1.10}
{\cal P}\left(t\right) \sim \exp\left(-C_d\ln^d t\right).
\end{equation}
Here $d$ is the dimensionality of the system and $t$ may have the meaning, e.g.,
of the current relaxation time, or the normalized local density of states at the
Fermi energy $\rho\left(E_F,{\bf r}\right)/\nu_d$, or the normalized square of the
local wave function amplitude $V\left|\psi_E\left({\bf r}\right)\right|^2$ of an
eigenstate with energy $E$. $\nu_d$ is the average $d$-dimensional density of 
states and $V$ is the volume of the sample. One notable exception is the
$e^{-\ln^2}$ law obtained in Ref. \cite{EfFal2} for the distribution of wavefunction
amplitudes in {\em three-dimensional} samples. We will discuss a possible origin
of this difference in Section \ref{disc}.

The coefficients $C_d$ in general depend on the
strength of disorder and, for $d=2$, on the sample size $L$. In the
two-dimensional case the calculations based on $\sigma$-models give $C_2=
\displaystyle{\frac{\beta\pi^2\nu_2D}{2\ln\frac{L}{l}}}$
\cite{AKL,EfFal1,EfFal2,Mirlin,Kh3}, where $p_F$ is the Fermi momentum,
$D=\displaystyle{\frac{1}{2}lv_F}$ is the electron diffusion constant, $l$ is the
mean free path (which is assumed to be much larger than the electron
wavelength $p_F^{-1}$), and $v_F$ is the Fermi velocity. $\beta$ is
a numerical coefficient which,  depending on the symmetry of
the ensemble of random potentials takes the following values: $\beta=1$ in
time-reversal-invariant systems (the orthogonal symmetry class), and $\beta=2$
when symmetry with respect to time reversal is completely broken (the  unitary
symmetry class). In Ref. \cite{Kh3} only the unitary ensemble was considered. We
will argue below that this value of $C_2$ can only be regarded as an
order-of-magnitude estimate.

In the one-dimensional case all eigenstates are localized, and the
distribution of wave function amplitudes has a simple Gaussian form (see below,
Eq.\ (\ref{eq3a.90})). Current relaxation times $\tilde{t}$, on the other hand, are
characterized by a much broader logarithmically normal distribution
\begin{equation}
\label{eq1.30}
{\cal P}\left(t\right)\sim\exp\left(-C_1\ln^2 \tilde{t}\right),
\end{equation}
where $C_1=l/2L$ for a sample of length $2L$ \cite{Kh3,AltPrig}.

In Refs.
\cite{EfFal2} and \cite{Mirlin} saddle-point solutions of the supersymmetric
nonlinear $\sigma$-models were obtained for the three-dimensional case. 
However, conventional nonlinear $\sigma$-models used in Refs.
\cite{AKL,Kh1,EfFal1,EfFal2,Mirlin} are low-``energy" effective field theories in
which the role of energy is played by the diffusion operator $D\nabla^2$. As such
they are only applicable to the description of phenomena that can be
characterized as diffusive -- the mean free path $l$ is the smallest length scale
in these theories. Formally this is expressed as a requirement that the scale of
spatial variations of the fundamental variables of the theory -- the Q-matrices
-- must be much larger than $l$.  The optimal fluctuations of the $Q$-matrices
computed in Refs. \cite{EfFal2,Mirlin} were found to vary  rapidly over distances
$\sim l$ in three-dimensional systems, which made it impossible to obtain
rigorous results. The coefficient $C_3$ was estimated in Ref. \cite{Mirlin} to be of
the order of $\left(p_Fl\right)^2$. A similar calculation in Ref. \cite{EfFal2} led to
the same estimate for $C_3$ but a {\em log-normal} rather than $e^{-C_3\ln^3t}$
functional form of the distribution function.

In an attempt to overcome the limitations of diffusive $\sigma$-models and
account exactly for the spatial variation of the  optimal fluctuation in 3D systems
at the scale of the mean free path, a generalized version of the model (ballistic
nonlinear $\sigma$-model) was introduced by Muzykantskii and Khmelnitskii in
Ref. \cite{Kh2} and used in Ref. \cite{Kh3} to compute the distribution of current
relaxation times. The functional form of Eq.\ (\ref{eq1.10}) was reproduced in that
calculation and a numerical value $\displaystyle{\frac{\pi}{9\sqrt{3}}
\left(p_Fl\right)^2}$ of $C_3$ was found. We will argue based on the calculations
presented below that all the results for the three-dimensional systems quoted
above contain serious errors.

In this article we propose an alternative method for the investigation of
distribution function asymptotics. Instead of integrating out the disorder
degrees of freedom and {\em then} looking for a saddle point of the resulting
effective field theory (the nonlinear $\sigma$-model), we suggest that
large-$t$ behavior of the distribution functions is governed by a saddle point
of the original theory in which the distribution functions are expressed as
functional integrals over both electronic and disorder degrees of freedom. In
other words, such a saddle point corresponds to an optimal fluctuation, i.e. the
highest-probability configuration of disorder that lets the electronic
eigenstates at a given energy $E$ have the desired property -- for example, an
anomalously large intensity $\,V\left|\psi_E\left({\bf r}\right)\right|^2$ at
some point ${\bf r}$. In our view, this approach possesses the advantage of
being applicable to systems of arbitrary dimensionality as well as to disorder
models more general than the commonly used model with a Gaussian
distribution of the potential fluctuations. It is also conceptually simpler and
makes the physical origin of the results much more transparent. In many
respects our approach is similar to the ideas utilized in Refs.
\cite{HalpLax,ZittartzLang}. We were able to reproduce the general $\ln {\cal P}
\sim -\ln^d t$ behavior of the distribution functions for $d=2$, $3$ as well as the
log-normal or Gaussian (depending on the precise definition of $t$) form of
${\cal P}$ for $d=1$. In two- and three-dimensional cases we obtain new values
for the coefficients $C_d$ correcting what we believe to be the errors in their 
previously published estimations. 

The rest of the paper is organized as follows. In Section \ref{sec2} we 
briefly describe the method and present the main results. A detailed
derivation of the basic equations and the analysis of the saddle point
solutions is deferred until Section \ref{sec3}. In Section \ref{disc}
we will compare our results with those obtained by employing nonlinear
$\sigma$-models.  A short summary and a list of open questions can be found in 
Section \ref{sec5}.

\section{Asymptotes of the Distribution of Wave Function Amplitudes.}
\label{sec2}

A typical wave function in a metallic sample is spread more or less uniformly
throughout the sample volume $V\sim L^d$ so that its amplitude does not differ
much from the average value of $1/\sqrt{V}$ anywhere in the sample. In the
metallic (or weakly localized) regime such states account for the bulk of the
distribution of the local density of states $\rho\left({\bf r}\right)$ or of the current
relaxation times \nolinebreak $\tilde{t}$. The chances of observing anomalously
large values of these quantities are related to the probability of finding an
``anomalously localized state". Such a state would be characterized by an
amplitude reaching a value much larger than the average at some point ${\bf r}$
inside the  sample.  In what follows we will concentrate on the distribution of
wave function amplitudes. Other distributions, such as those of current
relaxation times, local density of states, etc. can be more or less
straightforwardly derived within the framework of the same formalism. 

The problem can be formulated in the following way. Let us consider a spherical
(in 3D) or a disk-shaped (in 2D) conductor of radius $L$ and  compute the
probability that an eigenstate $\psi$ of energy $E$ (which we take to be equal to
the Fermi energy $E_F$) has an amplitude $\sqrt{t/V}$ in the center of the sample
(${\bf r}=0$), with $t\gg 1$ \cite{note1}. The distribution of the disorder potential is
assumed to be Gaussian,
\begin{equation}
\label{eq2.10}
{\cal W}\left[U\left({\bf r}\right)\right]={\cal N}_U\exp{\left(-\frac{\pi\nu_d\tau}{2}\int 
U^2\left({\bf r}\right)d{\bf r}\right)},
\end{equation}
where ${\cal N}_U$ is the normalization constant and $\tau$ is the mean free time.
Such a probability is then naturally expressed as 
\begin{equation}
\label{eq2.20}
{\cal P}\left(t\right)=\left\langle\delta\left(V\left|\psi\left(0\right)\right|^2-
t\right)\right\rangle_U,
\end{equation}
where $\langle\ldots\rangle_U$ denotes the averaging with the weight ${\cal W}$
over all possible configurations of $U$, and $\psi\left({\bf r}\right)$ is the solution
of the Schr\"{o}dinger equation with the Hamiltonian $H=H_0+U\left({\bf r}\right)$
and energy $E$. $H_0=\hat{\bf p}^2/(2m)$ is a Hamiltonian of free particles with a
mass $m$, and $\hat{\bf p}$ is the canonical momentum operator. 

The requirement that $E$ must be an eigenvalue of $H$ was enforced in Ref.
\cite{EfFal1} explicitly by introducing the corresponding $\delta$-function
into the definition of ${\cal P}$. In the alternative approach proposed here
-- the direct optimal fluctuation method -- this requirement is easier to
impose, if necessary, at a later stage in the calculations through appropriate
boundary conditions for the saddle point equations \cite{note3}.

Rewriting Eq.\ (\ref{eq2.20}) as a constrained functional integral and introducing
Lagrange multipliers to enforce the constraints (see Section \ref{sec3} for details)
one can demonstrate that the resulting ``action" ${\cal A}$ possesses a saddle
point, and the leading contribution to $\ln {\cal P}\left(t\right)$ is given by $-{\cal
A}_0$, the value of the ``action" at that point. With exponential accuracy the
results are:
\begin{mathletters}
\label{results}
\begin{equation}
\label{res3D}
{\cal P}\left(t\right)\sim\exp\left(-\kappa\left(p_F l\right)\ln^3 t\right),
\\\ \left(3D\right),
\end{equation}
\begin{equation}
\label{res2D}
{\cal P}\left(t\right)\sim\exp\left(-\pi^2\nu_2 D\frac{\ln^2 t}{\ln\left(L/r_0\right)}
\right), \\\
\left(2D\right),
\end{equation}
\end{mathletters}
where $\kappa$ is a number which is approximately equal to $3\times 10^{-3}$.
The distance $r_0$ in Eq.\ (\ref{res2D}) is of the order of the electron wavelength
$p_F^{-1}$. 

In both two- and three-dimensional cases, our answers have a
different $l$-dependence compared to the results obtained in Refs.
\cite{AKL,EfFal1,EfFal2,Mirlin,Kh3}. In the three-dimensional case $\ln{\cal P}$ is
proportional to the {\em first} power of $p_F l$ while all the results derived from
the nonlinear $\sigma$-model in three dimensions lead to a  $(p_F l)^2$
dependence. Similarly, in the two-dimensional case the logarithm in the
denominator contains $L/r_0$ instead of the $L/l$ established in Ref. \cite{AKL}
and reproduced with minor modifications in Refs. \cite{EfFal1,EfFal2,Mirlin,Kh3}. 
While in the two-dimensional case this difference leads only to small
corrections, the answer for a three-dimensional sample indicates a {\em
substantially increased} probability to observe an anomalously large eigenstate
amplitude as compared to the previous results. 

These asymptotes correspond to the optimal configurations of the potential 
$U\left({\bf r}\right)$ which are essentially Bragg mirrors, 
\begin{equation}
\label{eq2.40}
U\left({\bf r}\right)\propto \frac{\sin 2p_Fr}{r^{d-1}} 
\end{equation}
as $r\rightarrow\infty$. The $r\rightarrow 0$ behavior is more complicated. In one-
and two-dimensional systems the $r\rightarrow 0$ region does not contribute to
$\ln{\cal P}$ in the leading in $\ln t$ order. In three dimensions the shape of the
optimal fluctuation at small distances is a narrow potential well surrounded by a
wider potential barrier (Fig.\ \ref{fig1}) such that the whole structure supports a
narrow resonance in the $s$-wave channel. Anomalously large values of $t$ are
achieved by combining this local resonance with Bragg reflection at larger
distances.

The role of local resonances in producing large values of $t$ for any $d$ in the
interval $2\leq d<4$ can be illustrated by the following simplified version of the
direct optimal fluctuation method. Let us approximate the shape of the
resonance-producing potential as $U\left({\bf r}\right)=U_0\,\theta\left(r-a\right)
\theta\left(b-r\right)$, where
$\theta\left(r\right)$ is the step function, and $U_0$ ($>E_F$) and $b$ are
optimization parameters. The constant $a$ is determined from the condition that
there is only one resonance in the potential well and it has a given energy $E_F$ 
($a\sim p_F^{-1}=\left(2mE_F\right)^{-1/2}$).
 The wave function of the resonant state in such a potential is
characterized by exponential decay in the interval $a<r<b$. Assuming that the
amplitude of the wave function for $r>b$ is close to its average value $1/\sqrt{V}$
we can estimate $t\equiv V\psi^2\left(0\right)\sim\exp\left\{2\sqrt{2m
\left(U_0-E_F\right)}\left(b-a\right)\right\}$. Anticipating the final result
$b\sim p_F^{-1}\ln t$ we can neglect $a$ in the exponent. We then have 
\begin{equation}
\label{eq2.50}
U_0\approx E_F+\frac{1}{2m}\left(\frac{\ln t}{2b}\right)^2,
\end{equation}
and for $a\ll b$
\begin{equation}
\label{eq2.60}
{\cal A}=\frac{\pi\nu_d\tau}{2}\int U^2\left({\bf r}\right)d{\bf r}=
\frac{\pi\nu_d\tau}{2}V_dU_0^2b^d,
\end{equation}
where $V_d=\displaystyle{\frac{\pi^{d/2}}{\Gamma\left(d/2+1\right)}}$ is the
volume of a $d$-dimensional sphere of unit radius. Substituting Eq.\
(\ref{eq2.50}) into Eq.\ (\ref{eq2.60}) and minimizing with respect to $b$ we find
\begin{equation}
\label{eq2.65} 
b=p_F^{-1}\sqrt{\frac{4-d}{4d}}\ln t,
\end{equation}
and it follows then that $U_0=4E_F/\left(4-d\right)$. The resulting
estimate for the asymptotic behavior of the distribution function reads
\begin{mathletters}
\label{eq2.70}
\begin{eqnarray}
\label{eq2.71}
{\cal P}\left(t\right) & \sim & \exp\left\{ -  K_d\left(p_Fl\right)\ln^dt\right\}, \\
\label{eq2.72}
K_d & = & \frac{1}{2d}\left(\frac{1}{4\pi}\right)^{d-1}
\left(\frac{d}{4-d}\right)^{\frac{4-d}{2}}V_d^2.
\end{eqnarray}
\end{mathletters}
In the three-dimensional case the coefficient in front of $\ln^3 t$ is 
$\displaystyle{\frac{1}{2\cdot 3^{5/2}}\left(p_Fl\right)\approx
0.032\left(p_Fl\right)}$. It differs only by a number from a more
rigorous estimate quoted in Eq.\ (\ref{res2D}) suggesting that local resonances
play an important role in the three-dimensional  case. As it should be, the
coefficient $K_3\approx 0.032$ is {\em larger} than $\kappa$ since it would be
na\"{\i}ve to expect a box-shaped potential to be optimal.

In the two-dimensional case, large values of $t$ are not optimally produced by
local resonances: according to Eq.\ (\ref{res3D}) the coefficient in front of $\ln^2t$
is proportional to $\displaystyle{\left(\ln\frac{L}{r_0}\right)^{-1}}$, i.e. it depends
on the sample size.

The hypothesis that log-normal (in one and two dimensions) 
and $e^{-C_3\ln^3 t}$ (in 3D) asymptotics of the distribution functions may be 
produced by Bragg reflection locking the states in (or out of) certain regions of
the sample was put forward by Muzykantskii and Khmelnitskii in Ref. \cite{Kh3}.
They have explicitly demonstrated that a potential of the form 
$U\left(x\right)
=\left(p_F\ln\left(\tilde{t}\Delta\right)/mL\right)\cos\left(2p_Fx\right)$, where
$\Delta=\displaystyle{\frac{1}{\nu_dV}}$ is the mean level spacing,
traps an electron of energy $E_F$ in a one-dimensional 
sample of length $2L$ for time $\tilde{t}$. The probability of finding such a
configuration is
\begin{equation}
\label{eq2.80}
{\cal P}\left(\tilde{t}\right)\sim\exp\left(-\frac{l}{2L}
\ln^2 \left(\tilde{t}\Delta\right)\right),
\end{equation} 
which is the same as the log-normal long-time asymptotics of the exact solution
for the conductance \cite{AltPrig}. Our results confirm this hypothesis in
higher-dimensional samples. The most striking feature of our results is the fact
that a {\em single} configuration of the potential -- i.e. a single compact region of
the configurations space -- seems to be responsible for the large-$t$ tails of the
distribution functions. The alternative would be for a large number of more or
less equal-weight configurations to be able to produce the required wave
function amplitude $\displaystyle{\sqrt{\frac{t}{V}}}$, and the total probability
${\cal P}\left(t\right)$ would then be given by the sum of the weights of these
configurations. A strong argument against such an alternative is the fact that the
results obtained by simply singling out the most probable type of configurations
can lead to higher probabilities (significantly higher in the 3D case) than the
$\sigma$-model results which correspond to a sum over all configurations
\nolinebreak\cite{note2}.

\section{The Direct Optimal Fluctuation Method.}
\label{sec3}

In order to enforce the condition that $\psi$ in Eq.\ (\ref{eq2.20}) is a solution of
the Schrodinger equation we rewrite Eq.\ (\ref{eq2.20}) formally as
\begin{eqnarray}
\label{eq3.10}
{\cal P}\left(t\right) & = &
\int {\cal D}U\left({\bf r}\right){\cal W}\left[U\left({\bf r}\right)\right]
{\cal N}_{\psi}\left[U\right]\int {\cal D}\psi\left({\bf r}\right) \nonumber \\
& \times & \prod_{{\bf r}^{\prime}}
\delta\Bigl(\left(H-E\right)\psi\left({\bf r}^{\prime}\right)\Bigr)
\delta\left(V\left|\psi\left(0\right)\right|^2-t\right),
\end{eqnarray}
where ${\cal N}_{\psi}\left[U\right]$ is a $U$-dependent normalization constant
defined by
\begin{equation}
\label{eq3.20}
{\cal N}_{\psi}\left[U\right]\int {\cal D}\psi\left({\bf r}\right)\prod_{{\bf r}^{\prime}}
\delta\Bigl(\left(H-E\right)\psi\left({\bf r}^{\prime}\right)\Bigr)=1.
\end{equation}
Introducing auxiliary variables $\chi\left({\bf r}\right)$ and $\lambda$ associated
respectively with the first and the second $\delta$-functions in Eq.\ (\ref{eq3.10})
enables us to represent the distribution function as an unconstrained functional
integral:
\begin{mathletters}
\label{eq3.30}
\begin{eqnarray}
\label{eq3.31}
{\cal P}\left(t\right)= & {\cal N}_U & \int {\cal D}U\left({\bf r}\right)
{\cal N}_{\psi}\int {\cal D}\psi\left({\bf r}\right)
\int {\cal D}\left(\frac{\chi\left({\bf r}\right)}{2\pi}\right) \nonumber \\
& \times & \int \frac{d\lambda}{2\pi} e^{-{\cal A}\left[U,\psi,\chi,\lambda\right]}\, ,
\end{eqnarray}
\begin{eqnarray}
\label{eq3.32}
{\cal A}\left[U,\psi,\chi,\lambda\right] & = & \frac{\pi\nu_d\tau}{2}\int d{\bf r}\,
U^2\left({\bf r}\right) -i\lambda\left(V\left|\psi\left(0\right)\right|^2-t\right) 
\nonumber \\ 
 - i\int & d{\bf r} & \chi\left({\bf r}\right)\left(\frac{\hat{\bf p}^2}{2m} + U\left({\bf
r}\right) - E\right) \psi\left({\bf r}\right).
\end{eqnarray}
\end{mathletters}

The next step is to make a saddle-point approximation in Eq.\ (\ref{eq3.31})
utilizing the fact that $t$ is assumed to be large. The stationarity condition for
the action  Eq.\ (\ref{eq3.32}) leads to the following set of equations:
\begin{mathletters}
\label{eq3.40}
\begin{equation}
\label{eq3.41}
\pi\nu_d\tau U\left({\bf r}\right) - i\chi\left({\bf r}\right)\psi
\left({\bf r}\right)=0,
\end{equation}
\begin{equation}
\label{eq3.42}
\left(\frac{\hat{\bf p}^2}{2m}+U\left({\bf r}\right)-E\right)\psi
\left({\bf r}\right)=0,
\end{equation}
\begin{equation}
\label{eq3.43}
\left(\frac{\hat{\bf p}^2}{2m}+U\left({\bf r}\right)-E\right)\chi\left({\bf
r}\right)+ 2\lambda\sqrt{Vt}\delta\left({\bf r}\right)=0,
\end{equation}
\begin{equation}
\label{eq3.44}
V\left|\psi\left(0\right)\right|^2-t=0.
\end{equation}
\end{mathletters}

In deriving the saddle-point equations we neglect the $U$-dependence of the
normalization constant ${\cal N}_{\psi}$. The reasoning leading to this
approximation is explained in detail in the Appendix.

Below we will investigate only those solutions of Eqs.\ (\ref{eq3.40}) which
possess full rotational symmetry. While it is likely that even for
rotationally invariant boundary conditions there exist solutions which break
rotational symmetry we will assume that they are not optimal, i.e. that they
correspond to extremum points other than the global minimum of ${\cal A}$. The
mean free time enters these equations only through the $\pi\nu_d\tau$
factor in Eq.\ (\ref{eq3.41}). It can be absorbed into redefinitions of
$\chi\left({\bf r}\right)$ and $\lambda$, and therefore the above assumption
does not depend on the disorder strength.

Eliminating $\chi$ in favor of $U$ and $\psi$ and switching to dimensionless
variables $v\left(r\right)=\frac{U\left(\left|{\bf r}\right|\right)}{E_F}$,
$r=p_F\left|{\bf r}\right|$, we arrive (in the absence of magnetic field) at
the following system of equations ($\frac{d}{dr}\left(.\right)\equiv
\left(.\right)^{\prime}$):
\begin{mathletters}
\label{eq3.50}
\begin{equation}
\label{eq3.51}
y\left(r\right)=\ln^{\prime}\left(r^{\left(d-1\right)/2}\psi
\left(r\right)\right)
\end{equation}
\begin{equation}
\label{eq3.52}
y\,^{\prime}\left(r\right)+y^2\left(r\right)+\frac{1}{4r^2}\delta_{d,2}+1=
v\left(r\right),
\end{equation}
\begin{equation}
\label{eq3.53}
2y\left(r\right)\left(r^{d-1}v\left(r\right)\right)- 
\left(r^{d-1}v\left(r\right)\right)^{\prime}=\overline{\lambda},
\end{equation}
\end{mathletters}
where 
\begin{equation}
\label{eq3.54}
\overline{\lambda}=\frac{2i\lambda t}{\pi\nu_d\tau}\frac{p_F^d}{E_F^2S_d},
\end{equation}
$\delta_{d,2}$ is the Kronecker symbol and $S_d=dV_d$ is the area of a
$d$-dimensional sphere of unit radius. Equations (\ref{eq3.51}) and (\ref{eq3.52})
together are equivalent to the Schr\"{o}dinger equation for the radial part of the
wave function, while Eq.\ (\ref{eq3.53}) provides the nontrivial connection
between the wave function and the potential that is necessary to achieve a given
value of $t$ at a minimal ``cost" in terms of the weight function ${\cal W}$.
$\lambda$ can be viewed as a parameter effecting a Legendre transformation of
${\cal A}$, so $\left(\lambda,t\right)$ -- alternatively, 
$\left(\overline{\lambda},\ln t\right)$ -- form a pair of conjugate variables.

It follows from Eq.\ (\ref{eq3.53}) that $\overline{\lambda}$ must be real, so the
saddle-point value of $\lambda$ is purely imaginary. The saddle-point action
${\cal A}_0$ is expressed in terms of the solution $v\left(r\right)$ of  
Eqs.\ (\ref{eq3.50}) as 
\begin{equation}
\label{eq3.55}
{\cal A}_0=\frac{\pi\nu\tau}{2}\frac{E_F^2}{p_F^d}S_d
\int_0^L r^{d-1}v^2\left(r\right)dr.
\end{equation}

The choice of  boundary conditions to Eqs.\ (\ref{eq3.50}) is not entirely
straightforward, and we would like to motivate it by appealing to an analogy with
ordinary one-dimensional Schr\"{o}dinger equation. In the variational formulation
of quantum mechanics the Schr\"{o}dinger equation appears as a stationarity
condition of a functional
\begin{equation}
\label{eq3.56}
{\cal F}=\int_0^L dx\psi^{*}\left(x\right)\left(\hat{H}-E\right)\psi\left(x\right),
\end{equation}
where $\hat{H}$ is the Hamiltonian, $E$ is the particle energy, and $\psi$ must
satisfy two boundary conditions (e.g., $\psi\left(0\right)=\psi\left(L\right)=0$) and
also the normalization condition $\int_0^L dx |\psi|^2=1$. The overall number of
conditions for this second-order equation is three, which makes it over-defined.
As a result, solutions exist only for a discrete set of values of $E$. One of these
values would correspond to the ground state and thus to an absolute minimum
of ${\cal F}$, while the rest would simply correspond to its stationary points. In
the $L\rightarrow\infty$ limit the set of allowed $E$ values becomes infinitely
dense so that $E$ effectively becomes a continuous variable. 

An alternative way to look at the Schr\"{o}dinger equation (often useful in
numerical calculations  \cite{NumRec}) is to consider $E$ as an extra unknown
function satisfying a trivial equation
\begin{equation}
\label{eq3.57}
\frac{d}{dx}E=0. 
\end{equation}
One then has effectively three first-order equations with three extra conditions
imposed on them. However, now these equations are nonlinear, and the three
conditions do not specify the solution uniquely. Rather, there exists a family of
solutions corresponding to various quantum states of the system, one of which
minimizes the functional ${\cal F}$. Again, in the $L\rightarrow\infty$ limit the
family of solutions becomes continuous.

Returning to the problem of specifying a set of conditions for Eqs.\ (\ref{eq3.50})
we see that equation (\ref{eq3.53}) is a generalization of Eq.\ (\ref{eq3.57}).
$\overline{\lambda}$ plays the role of an eigenvalue, so we can formally
supplement Eqs.\ (\ref{eq3.50}) with $\overline{\lambda}\,^{\prime}=0$, increasing
the number of equations to four.  The following set of conditions must be
imposed on this system of equations.

First, a boundary condition is provided by Eq.\ (\ref{eq3.44}):
\begin{mathletters}
\label{eq3.60}
\begin{equation}
\label{eq3.61}
\psi\left(0\right)=\sqrt{\frac{t}{V}}.
\end{equation}
The normalization requirement
\begin{equation}
\label{eq3.62}
S_d\int_0^L r^{d-1} dr \psi^2\left(r\right)=1
\end{equation}
provides the second condition.

Another boundary condition is derived from regularity requirements on
$v\left(r\right)$. Since $v\left(r\right)$ can have at most a square integrable
singularity as $r\rightarrow 0$, it follows that in three dimensions
\begin{equation}
\label{eq3.63}
\lim_{r\rightarrow 0}\left(\sqrt{r}\psi^{\prime}\left(r\right)\right)=0.
\end{equation}
In two dimensions we simply have a condition that $\psi^{\prime}$ must be finite
at $r=0$, and for a one-dimensional system the corresponding requirement is 
$\psi^{\prime}\left(0\right)=0$. Finally, a fourth condition may be provided by
another boundary condition on $\psi$, e.g.
\begin{equation}
\label{eq3.64}
\psi\left(L\right)=0. 
\end{equation}
\end{mathletters}
However the problem of minimizing ${\cal A}$ is not solved uniquely by the set
of equations (\ref{eq3.50}) together with the conditions (\ref{eq3.60}). As in
the ordinary quantum mechanical case, we are faced with a family of solutions,
only one of which actually corresponds to the absolute minimum ${\cal A}_0$ of
the action ${\cal A}$. Finding this ``ground state" solution requires a
parametrization of the whole discrete family of solutions which presents a
difficult task. It can nevertheless be somewhat simplified by employing the
following observation. For large enough $L$ the set of solutions of Eqs.\
(\ref{eq3.50}) may be approximately considered continuous. It is then possible
to replace the condition (\ref{eq3.64}) with a free boundary condition. The
result will be that the family of allowed solutions will now admit a
continuous parametrization, simplifying the task of determining the minimum of
the functional ${\cal A}$. In practice this parametrization depends on the
dimensionality of the sample, and we will consider the cases of $d=1$, $2$ and
$3$ separately. Employing a continuous parametrization which effectively
corresponds to relaxing the boundary condition Eq.\ (\ref{eq3.64}) means not
enforcing the condition that $E_F$ is an eigenvalue of the Hamiltonian. As was
already mentioned \cite{note3} such an approximation leads only to errors that
are beyond the exponential accuracy with which asymptotes of ${\cal
P}\left(t\right)$ are determined in the optimal fluctuation method and which can
therefore be safely neglected.

\subsection{One-dimensional wire.}
\label{sec3a}

As a ``toy" model, we consider first a purely one-dimensional
disordered wire of length $2L$.  All the eigenstates in this case are
exponentially localized with a localization length being of the order of the mean
free path $l$. A typical amplitude of such a state in its domain of localization is
$1/\sqrt{l}\gg 1/\sqrt{2L}$, so the optimal fluctuation method cannot be applied
to the computation of the distribution function ${\cal P}\left(t\right)$ for
$t\lesssim L/l$. Amplitudes larger than that are due to the states that are
characterized by an anomalously small localization length and the probability of
finding such states is determined by the probability of the corresponding optimal
fluctuation.

The system of Eqs.\ (\ref{eq3.50}) takes the form
\begin{eqnarray}
\label{eq3a.10}
& & v\left(x\right)  =  1+y^2\left(x\right)+{y\,}^{\prime}\left(x\right), \nonumber \\
& & 2y\left(x\right)v\left(x\right)-{v\,}^{\prime}\left(x\right)=
\frac{1}{2}\overline{\lambda}\,
\mbox{\text Sgn}\,x,
\end{eqnarray}
where $x$ is the dimensionless coordinate along the wire, and primes denote
differentiation with  respect to \nolinebreak $x$. Note that when $d=1$, $y$ is
simply the logarithmic derivative of the wave function. Eliminating $v\left(x\right)$
and integrating once, we arrive (for $x>0$) at
\begin{equation}
\label{eq3a.20}
\left({y\,}^{\prime}\right)^2=\left(y^2+1\right)^2+\overline{\lambda}\,y+C.
\end{equation}
Here $C$ is a constant of integration which parametrizes the stationary points of
${\cal A}$ as described at the end of the preceding subsection. Its value is fixed
by a minimization procedure which is formally expressed as 
\begin{equation}
\label{eq3a.25}
{\cal A}_0=\min_C {\cal A}\left(C\right)
\end{equation}

Although Eq.\ (\ref{eq3a.20}) is
exactly integrable in terms of elliptic functions, it is sufficient to make a
perturbative expansion in $\overline{\lambda}$ and $C$ in order to recover the
leading order dependence on $t$. Rewriting Eq.\ (\ref{eq3a.20}) as
\begin{equation}
\label{eq3a.30}
dx=-\frac{dy}{R\left(y\right)},\,\,\,\,\,\,\,\,\,R\left(y\right)=\sqrt{\left(y^2+1\right)^2+
\overline{\lambda}\,y+C},
\end{equation}
we obtain the period $T$ of the wave function oscillations
\begin{equation}
\label{eq3a.40}
T=2\int_{-\infty}^{\infty}\frac{dy}{R\left(y\right)}
\end{equation}
and the logarithm of the ratio of the wave function amplitudes at $x=0$ and $x=L$:
\begin{equation}
\label{eq3a.50}
\ln\frac{\psi\left(0\right)}{\psi\left(L\right)}=2\frac{L}{T}\int_{-\infty}^{\infty}
\frac{ydy}{R\left(y\right)}.
\end{equation}
We assume that the length of the sample corresponds to an integer number of
periods, so that if $y\left(0\right)=0$ (from symmetry arguments) it follows
also that $y\left(-L\right)=y\left(L\right)=0$. The relative error introduced into
Eq.\ (\ref{eq3a.50}) by neglecting the fractional part of $L/T$ is
$O\left(1/L\right)$.

It is convenient to use the logarithm in the l.h.s. of Eq.\ (\ref{eq3a.50}),
which we will denote as $\ln \theta$, as a parameter of the distribution
function. It can be easily seen that the contribution to ${\cal P}$ of the
Jacobian of the transformation from $\theta$ to $t$ is only a prefactor
which does not change the leading order term in the exponent. Expanding the
r.h.s. of Eqs.\ (\ref{eq3a.40}) and (\ref{eq3a.50}) in both
$\overline{\lambda}$ and $C$ we obtain
$-\pi\overline{\lambda}/16=\left(T/2L\right)\ln\theta$ and 
$T=2\pi+O\left(C\right)$. For the action ${\cal A}$ we then find to the
lowest order (with $S_d=2$)
\begin{equation}
\label{eq3a.60}
{\cal A}=\pi\nu_1\tau\frac{E_F^2}{p_F}\frac{2\pi L}{32T}
\left(\overline{\lambda}^2+3C^2\right).
\end{equation}
It is thus obvious that, at least to the lowest order in $1/L$, the minimum of
${\cal A}$ is achieved when $C=0$. Combining the results together we have for
the asymptotic tail of the distribution  function with exponential accuracy
\begin{equation}
\label{eq3a.70}
{\cal P}\left(\theta\right)\sim\exp\left\{-\frac{2l}{L}\ln ^2\theta\right\}.
\end{equation}
From Eq.\ (\ref{eq3a.30}) and the fact that the 
change in the wave function amplitude over one period (increase for $x<0$ or
decrease for $x>0$) does not depend on $x$, we deduce the
exponential form of the wave function envelope
\begin{equation}
\label{eq3a.80}
\psi\left(x\right)\propto e^{-\frac{2\left|x\right|}{L}\ln\theta}.
\end{equation}
Normalizing the wave function to $1$ and using $V=2L$ we obtain
$t=2L\psi^2\left(0\right)=2\left(\ln\theta\right)$. We thus finally
arrive at
\begin{equation}
\label{eq3a.90}
{\cal P}\left(t\right)\sim\exp\left\{-\frac{l}{2L}t^2\right\}.
\end{equation}
As explained at the beginning of this subsection the applicability of the above
formula is restricted to the region $L/l<t <p_FL$ (the second inequality
ensures the validity of the expansion in $\overline{\lambda}$ and $C$). 
As required, the corresponding value of the localization length $L/\ln\theta$ is
much smaller than $l$.

This model calculation can be used to illustrate the following points. First,
to the lowest order in $\overline{\lambda}$ we have
\begin{equation}
\label{eq3a.100}
y\left(x\right)=\cot\left(x+\varphi\right)+\frac{\overline{\lambda}}{4} 
\sin^2\left(x+\varphi\right),
\end{equation}
where $\varphi$ is a phase shift, and 
\begin{equation}
\label{eq3a.110}
v\left(x\right)=\frac{1}{2}\overline{\lambda}\sin 2\left(x+\varphi\right), 
\end{equation}
so the shape of the optimal configuration of the potential is indeed a Bragg
mirror. 

Second, up to a constant, $\theta ^2/\Delta$ can be identified with
$\tilde{t}$ -- the delay time in the electric response of an open sample at $E_F$ 
\cite{Kh3}, and the distribution in Eq.\ (\ref{eq3a.70}) is seen to coincide with the
exact answer of Altshuler and Prigodin \cite{AltPrig} quoted at the end of
Section \ref{sec2}. Third, the sign of $\overline{\lambda}$ is opposite to
that of $\ln\theta$, so negative values of $\overline{\lambda}$ correspond to
the wave function amplitude decreasing from the center of the sample outwards.
The same will hold true in two and three dimensions. The amplitude of the
oscillating potential is constant throughout the sample. A similar observation
was made in Ref. \cite{Kh3} -- in a one-dimensional wire the gradients of the
supersymmetric density matrix corresponding to the optimal solution were found
to be $x$-independent. 

\subsection{Two-dimensional conductors}
\label{sec3b}

In dimensions higher than one the system (\ref{eq3.50}) cannot be integrated
exactly in terms of the standard elementary or special functions. Nevertheless,
numerical methods combined with the asymptotic analysis allow us to
fully investigate the behavior of its solutions in various regimes. In 2
dimensions the solution can be obtained as an asymptotic expansion in
$\overline{\lambda}/r$. The values of $t$ for which this expansion breaks down
turn out to lie close to the limiting value $\pi L^2$, and are thus not very
interesting. The leading terms of the expansion of $v\left(r\right)$ are
\begin{equation}
\label{eq3c.10}
v\left(r\right)\approx\frac{\overline{\lambda}}{2r}\sin\left(2r+2\varphi\right)
+ \frac{\eta}{r}\sin^2\left(r+\varphi\right).
\end{equation}
In this expression $\eta$ is a constant of integration which plays a role
analogous to the role of $C$ in the one-dimensional case -- that of a
minimization parameter. Just as in the one-dimensional case, the minimum of
${\cal A}$ is reached when $\eta=0$. We will give a perturbative proof of this
statement at the end of this subsection. With $\eta=0$ the asymptotic
expansion of $y\left(r\right)$ has the form
\begin{equation}
\label{eq3c.15}
y\left(r\right)\approx\cot\left(r+\varphi\right)+
\frac{\overline{\lambda}}{4r}\sin^2\left(r+\varphi\right).
\end{equation}

It is convenient to define the parameter $\theta$ as
\begin{equation}
\label{eq3c.17}
\theta =\frac{\psi\left(0\right)}{\sqrt{L}\psi\left(L\right)}
\end{equation}
in order to cancel out the overall
$1/\sqrt{r}$ dependence of the wave function. $\theta$ is always large in the
asymptotic region $t\gg 1$.
Integrating $y\left(r\right)$ in the sense of the principal value we obtain
\begin{equation}
\label{eq3c.20}
\ln\theta=-\mbox{\text P}\int_{r_0}^Ly\left(r\right)dr+\ln\theta_1
\approx\frac{|\overline{\lambda}|}{8}\ln
\frac{L}{r_0}+\ln\theta_1,
\end{equation}
where the length scale $r_0$ is defined as
$r_0\sim\max\,\left(1,|\overline{\lambda}|\right)$, and  $\ln\theta_1$
represents the contribution to the integral from distances $r\lesssim r_0$.
When $|\overline{\lambda}|\lesssim1$ this contribution can be neglected with
logarithmic accuracy. Inverting Eq.\ (\ref{eq3c.20}) we obtain
$|\overline{\lambda}|=8\ln\theta /\ln\displaystyle{\frac{L}{r_0}}$. The
dimensionless integral in the saddle point action $\int_0^Lv^2rdr$ evaluates
to $8\ln^2\theta/\ln\displaystyle{\frac{L}{r_0}}$, and we find
\begin{equation}
\label{eq3c.30}
{\cal P}\left(\theta\right)\sim\exp\left\{-4\pi^2\nu_2D
\frac{\ln^2\theta}{\ln\left(L/r_0\right)}\right\}.
\end{equation}

The envelope of the wave function corresponding to the solution described by
Eq.\ (\ref{eq3c.10}) is
\begin{equation}
\label{eq3c.35}
\psi\left(r\right)\sim\frac{\psi\left(0\right)}{\sqrt{r}}
\left(r_0/r\right)^{\ln\theta/\ln\frac{L}{r_0}},
\end{equation}
where we have approximated
$\sqrt{r_0}\psi\left(r_0\right)\approx\psi\left(0\right)$. For $r_0\sim1$ the
error introduced by this approximation leads only to $O\left(1\right)$
corrections to the large logarithm $\ln t$. The normalization integral
\begin{equation}
\label{eq3c.40}
N=2\pi\psi^2\left(0\right)\int_{r_0}^Ldr\left(\frac{r_0}{r}\right)^{2\ln\theta
\ln \frac{L}{r_0}}
\end{equation}
has two distinct regimes. When $\theta^2$ is greater than $L/r_0$ the integral
is dominated by small distances, and the dependence of $t$ on $\theta$ is
weak. We will not be considering this regime here because it corresponds to
wave function amplitudes that are close to the limiting value $t=\pi L^2$. In the
opposite case we have $N=2\pi\psi^2\left(0\right)r_0^\alpha
L^{1-\alpha}/\left(1-\alpha\right)$, where
$\alpha=\ln\theta^2/\ln\frac{L}{r_0}$. Up to irrelevant constants we then have
$\ln t\approx\ln\theta^2$, and the distribution function becomes
\begin{equation}
\label{eq3c.50}
{\cal P}\left(t\right)\sim\exp\left\{-\pi\nu_2D
\frac{\ln^2 t}{\ln\left(L/r_0\right)}\right\}.
\end{equation}

This result has essentially the same form as the one obtained by Falko and
Efetov \cite{EfFal2} using the $\sigma$-model formalism. It differs, however, from
their result in two important aspects.
First, the logarithm in the denominator is cut off
at distances $\sim 1$ ($p_F^{-1}$ in conventional units) rather than at $r\sim
l$, as in Ref. \cite{EfFal2}. We believe that this difference simply reflects
the fact that both the approach employed here and the saddle-point solution of
the $\sigma$-model represent {\em the same} saddle point of the underlying
theory. However, in order to describe this saddle point exactly one has to
include short-wavelength degrees of freedom which are lost in the derivation
leading to the $\sigma$-model, and that accounts for its inability to obtain
the correct cut-off scale.

The second difference may turn out to be an indication of a deeper problem with
the $\sigma$-model. Our calculation is performed explicitly for the case when
there are no perturbations breaking the symmetry with respect to time reversal.
Nevertheless, the numerical coefficient in the exponent coincides with the
answer obtained in Ref. \cite{EfFal2} for the {\em unitary} case ($\beta=2$), when
the symmetry with respect to time reversal is completely broken. In order to see
what effect violation of time-reversal invariance would have in our approach, we
explicitly introduce magnetic field into the system of equations (\ref{eq3.40}).
Choosing the direction of the field ${\bf H}$ along the  $\hat{z}$ axis
(perpendicular to the two-dimensional sample) and writing ${\bf
A}=\frac{1}{2}{\bf H}\times {\bf r}$ we can immediately see that under the
assumption of circular symmetry of the solution $\psi$, only the terms
quadratic in ${\bf A}$ survive. For weak enough fields $HL^2\ll\phi_0$ (where
$\phi_0$ is the magnetic flux quantum) these terms can be neglected because
they only lead to exponential decay of the wave function on a scale larger
than the sample size. Thus the set of equations\ (\ref{eq3.50}) is unchanged.
On the other hand, the fields of this magnitude are sufficient to suppress the
Cooperon contribution in the $\sigma$-model calculation leading to a crossover
between the orthogonal and unitary symmetry classes. Therefore our results
reproduce {\em exactly} the $\sigma$-model answer in the case of broken
time-reversal invariance (the unitary symmetry class), but they predict a {\em
faster} decay of the distribution function at large $t$ in the orthogonal
case. If the fault lies with our calculation, it would then mean that while in
the unitary case a single optimal fluctuation of the potential is responsible
for the high-amplitude tail of the distribution of the wave function intensities,
an exponentially large number of non-circularly-symmetric configurations
would have to contribute to the probability of finding large wave function
amplitudes in the orthogonal case. (It is certainly {\em very} unlikely that
accounting for the fluctuations around the saddle point defined by Eq.\
(\ref{eq3.40}) can change the coefficient of a $\ln^2 t$ term in the
exponential). While this possibility cannot be completely discounted without a
more detailed study of the system Eqs.\ (\ref{eq3.40}), we would like to
propose an alternative explanation. 

The expansion into the sum over quasiclassical paths that underlies the
nonlinear $\sigma$-model, and the subsequent separation into the diffuson and
Cooperon contributions implies that self-retracing paths constitute a set of
measure zero in the overall sum. This is certainly a rigorous assumption when
there is no ``intrinsic" mechanism in the calculation that would strongly
favor some types of quasiclassical paths over others -- as is almost always the
case in clean chaotic systems where the zero mode treatment of the
$\sigma$-model is applicable. However, when a saddle point approximation is
being made in order to compute the probability of some rarely occurring event,
one effectively selects certain members of the ensemble of random potentials
that are favorable for such an event. Within the $\sigma$-model formalism this
selection occurrs as a restriction on the possible configurations of the
effective order parameter (the $Q$-matrix, or the supersymmetric density
matrix $g$ in the formalism of Muzykantskii and Khmelnitskii \cite{Kh2,Kh3}),
and the information about the actual configurations of the potential that are
being selected is lost. On the other hand, as indicated by the calculation
presented here, these configurations are rather special. Indeed, one can
interprete the appearance of a quasilocalized state possessing an anomalously
large amplitude at some point ${\bf r}$ as an opening of a quasi-gap in the
spectrum of $s$-waves with respect to ${\bf r}$. In a weak periodic potential
emergence of a gap can already be seen in the second order of the perturbation
theory, and the second order terms in the expansion into a sum over
quasiclassical paths by necessity contain only self-retracing trajectories.
Such contributions are then overcounted by the $\sigma$-model when it takes
into account the same self-retracing path twice -- as a diffuson {\em and} as a
Cooperon.

While the above argument is rather speculative, it nevertheless presents a
picture which, if confirmed, would restrict the applicability of saddle-point
solutions of the nonlinear $\sigma$-model. We believe that this issue deserves
further consideration.

To complete the derivation of the distribution function presented in this
subsection, we outline the proof of the statement that $\eta=0$ is the optimal
choice. The solution $\psi$ of equations (\ref{eq3.40}) can be written as
$\psi\left(r\right)=
\frac{A\left(r\right)}{\sqrt{r}}\sin\left(r+\varphi\left(r\right)\right)$. 
Unlike the $\eta=0$ case, the phase $\varphi$ does not have a finite limit as
$r\rightarrow\infty$. The logarithmic derivative $\tilde{y}\left(r\right)$ of
the amplitude function $A\left(r\right)$ can be expanded in an asymptotic
Fourier series of the form
\begin{eqnarray}
\label{eq3c.60}
\tilde{y}\left(r\right) & \sim &\sum_{n=1}^{\infty}\frac{1}{r^n}
\left\{y_{\left(n,0\right)} 
 +  \sum_{m=1}^{\infty}\left[y_{\left(n,m\right)}^{\left(c\right)}
\cos m\left(r+\varphi\left(r\right)
\right) \right. \right. \nonumber \\
& + & \left. \left. y_{\left(n,m\right)}^{\left(s\right)}
\sin m\left(r+\varphi\left(r\right)\right)\right]\right\},
\end{eqnarray}
and $\varphi^{\prime}\left(r\right)$ can be represented in a similar way.
Substituting these expansions into Eqs.\ (\ref{eq3.40}) we find that
$y_{\left(1,0\right)}=\overline{\lambda}/8$. Thus the choice of the constant of
integration $\eta$ does not affect the relation between $y_{\left(1,0\right)}$ and
$\overline{\lambda}$. Integrating $y\left(r\right)$ over $r$ we obtain
$y_{\left(1,0\right)}\ln\frac{L}{r_0}$ and therefore the relation between $\ln\theta$
and $\overline{\lambda}$ established in Eq.\ (\ref{eq3c.20}) is also
$\eta$-independent to the leading order in $\overline{\lambda}$. On the other
hand, adding the term $\frac{\eta}{r}\sin^2\left(r+\varphi\right)$ to $v$ can only
increase the value of the integral $\int_0^{L}v^2 rdr$ because the cross-term in
the expansion of the square integrates to zero. It follows then that the minimum
of the integral for a given value of $\theta$ is achieved by setting $\eta=0$. Note,
however, that, as in the one-dimensional case, this proof is perturbative -- it
relies on the possibility to expand the solutions in powers of 
$\overline{\lambda}/r$.

\subsection{The three-dimensional case.}
\label{sec3c}

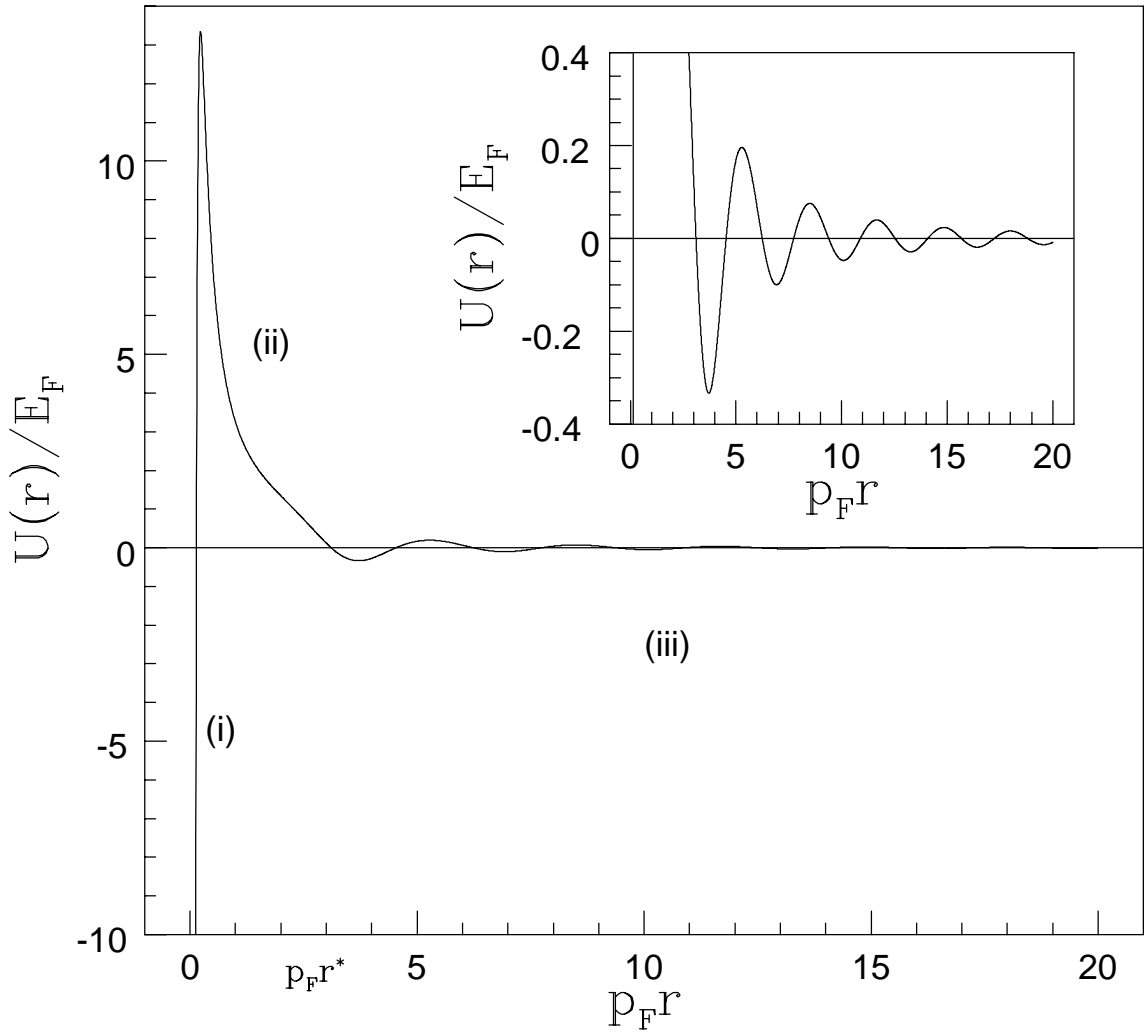
\begin{figure}
\caption{The profile of an optimal configuration of the potential in three
dimensions as a function of the radial coordinate.}
\label{fig1}
\end{figure}

In the three-dimensional case, it can be self-consist-ently demonstrated that
large values of $t$ correspond to large negative values of $\overline{\lambda}$.
As a result, there exists a range of values of $r$ where expansion in
$\overline{\lambda}/r^{d-1}=\overline{\lambda}/r^2$ is impossible.  A typical
solution which was obtained numerically using the so-called relaxation method
\cite{NumRec} is shown in Fig.\ \ref{fig1}. One can distinguish three
asymptotic regimes: (i) $r\ll r_0$, (ii) $r_0\ll r\ll r^*$, and (iii) $r\gg r^*$; it will be
shown below that $r_0=1/|\overline{\lambda}|$ and
$r^*=\sqrt{|\overline{\lambda}|/2}$. The first
region corresponds to a potential well, the second one -- to a potential barrier. 
Taken together, these two regions support a resonance in the $s$-wave
channel at the energy $E_F$. However, besides a potential-well --
potential-barrier  combination, resonant scattering can be also caused by a weak
periodic potential (Bragg reflection), and that is exactly what the third region
corresponds to. An interesting consequence of the solution presented in Fig.\
\nolinebreak\ref{fig1} is that the optimal way to achieve large values of $t$ in
three dimensions is to combine the two effects in the ``right" proportion. 

Analytically the three regions in Fig.\ \ref{fig1} are described by the following
asymptotic formulae. In the first region, $v\left(r\right)$ diverges as
\begin{equation}
\label{eq3b.05}
v_1\left(r\right)\sim\overline{\lambda}/r+\overline{\lambda}^2
\ln r+\left(1+c+\frac{7}{12}\overline{\lambda}^2\right),
\end{equation}
where $c$ is an arbitrary constant which
cannot be determined from the boundary conditions Eq.\ (\ref{eq3.60}). Note
that the singularity is weak enough so that it produces only a finite contribution
to the saddle-point action ${\cal A}_0$. The behavior of the wave function in this
regime is given by 
\begin{equation}
\label{eq3b.10}
y\left(r\right)\approx
\frac{1}{r}+\frac{1}{2}\overline{\lambda}+\frac{1}{3}\overline{\lambda}^2 r\ln r+c\,r,
\end{equation}
which corresponds to
\[\psi\left(r\right)\approx\psi\left(0\right)\left(1+\overline{\lambda}r/2+ 
O\left(r^2\ln r\right)\right). 
\]
The expansion in Eq.\ (\ref{eq3b.05}) breaks down for
$r\sim 1/|\overline{\lambda}|$ which gives the approximate value for $r_0$.
In the second
region the solution can be obtained with the help of the quasiclassical
approximation, $y^2\left(r\right)\approx v\left(r\right)$, and the result is 
\begin{equation}
\label{eq3b.15}
v_2\left(r\right)\approx
\left(\frac{|\overline{\lambda}|}{2r^2}\right)^{2/3}.
\end{equation}
Finally, in the third region, where the
$\overline{\lambda}/r^2$ expansion works,
$v\left(r\right)$ has the asymptotic form
\begin{equation}
\label{eq3b.20}
v_3\left(r\right)\sim\frac{\overline{\lambda}}{2}\frac{\sin\left(2r+2\varphi\right)}{r^2}
+ \eta\frac{\sin^2\left(r+\varphi\right)}{r^2}+O\left(\frac{1}{r^3}\right).
\end{equation}
The constants $\eta$ and $\varphi$ in this expression are analogous to their
counterparts in the two-dimensional case. The phase variable $\varphi$ again
has the meaning of the wave function phase shift; it is finite due to the rapid
decay of the potential at large $r$ irrespective of the value of $\eta$. Constants
$\eta$ and $\varphi$ are not independent: they both can be regarded as functions
of $c$. Either $\eta$ or $c$ can be chosen as a minimization parameter.

We have not been able to establish the analytical dependence $\eta\left(c\right)$;
however, based on the numerical analysis the qualitative features of this
dependence can be described as follows. For a given $\overline{\lambda}$ there
exists a critical value $c_0\left(\overline{\lambda}\right)$ such that if $c>c_0$, the
third (oscillatory) region never develops. Instead, $v\left(r\right)$ exhibits a
singularity at some finite value of $r$, leading to a divergent integral in ${\cal
A}_0$. So the values $c>c_0$ correspond to unphysical solutions. Exactly at the
critical value $c_0$ the oscillations are also absent, and
$v\left(r\rightarrow\infty\right)\approx 1+\overline{\lambda}^2/(4r^4)$. For slightly
smaller $c$, $\eta\left(c\right)$ is large and positive ($\gg |\overline{\lambda}|$),
and the oscillations appear only after a more or less protracted intermidiate
regime in which $v\left(r\right)\approx 1$. An estimate of the point
$r^*$ at which the onset of the oscillatory behavior occurrs can be obtained by
noting that the oscillations of $v\left(r\right)$ are driven (through Eq.\
(\ref{eq3.50})) by oscillations of the wave function. Therefore the amplitude of the
oscillations in Eq.\ (\ref{eq3b.20})  cannot significantly exceed $1$ so as to
preserve the oscillatory -- rather than exponentially damped -- behavior of
the wave function. This requirement leads to 
\begin{equation}
\label{eq3b.25}
r^*\approx\frac{1}{2}\left(\overline{\lambda}^2+\eta^2\right)^{1/4}.
\end{equation}
As $c$ decreases further,
$\eta$ monotonically decreases as well, eventually covering the whole
$(+\infty,-\infty)$ interval. 

When $\eta$ is positive or small negative ($\eta\gtrsim -|\overline{\lambda}|$) the
behavior of $v\left(r\right)$ and $y\left(r\right)$ in the regions (i) and (ii) depends
on $\eta$ (or $c$) only very weakly so that this dependence can be ignored in
computing the contribution of these regions to ${\cal A}$. Larger negative values
of $\eta$ start affecting the length of region (ii), and may lead to an emergence of
a hybrid regime in which $v$ oscillates non-harmonically with an $r^{-4/3}$
envelope. 

In what follows we will assume that similarly to the two-dimensional case the
minimal value ${\cal A}_0$ is achieved by setting $\eta =0$, even though the
breakdown of the asymptotic expansion in
$\displaystyle{\frac{\overline{\lambda}}{r^2}}$ at small distances makes it
impossible to construct an analytical proof of this statement analogous to the
proofs for one- and two-dimensional systems. This assumption is borne out by
numerical analysis. It is certainly obvious that positive values of $\eta$ can never
be optimal because of a corresponding ``costly" $v\sim 1$ region. As for large
negative values, they are ruled out by the fact that a pronounced ``hybrid" regime
never appears in numerical solutions. Setting $\eta=0$ reduces Eq.\
(\ref{eq3b.25}) to $r^*\approx \sqrt{|\overline{\lambda}|/2}$. 

With $\eta\approx 0$, both $v_2\left(r\right)$ and $v_3\left(r\right)$ reach values
$\sim 1$ at $r=r^*$. While approximation
schemes devised for $v\gg 1$ and $v\ll 1$ break down around $r^*$, we can
calculate contributions $\theta_1$ and $\theta_2$ to
$\theta=\frac{\psi\left(0\right)}{L\psi\left(L\right)}=\theta_1\theta_2$ from regions
$r<r^*$ (i-ii) and $r>r^*$ (iii) separately. Similar to the two-dimensional case,
the  $L$ factor in the denominator of $\theta$ is introduced to cancel the overall
$1/r$ dependence of the wave functions, which is an artefact of spherically
symmetric boundary conditions. 

(i) - (ii) We combine together the first and the second regions because the wave
function amplitude does not change appreciably in the very short region (i).
Region (ii) corresponds to stretched exponential decay of $\psi$. Integrating
$y\left(r\right)=-\sqrt{v_2\left(r\right)}$ over $r$ from $1/|\overline{\lambda}|$ to
$r^*\approx\sqrt{|\overline{\lambda}|/2}$ we obtain
\begin{equation}
\label{eq3b.30}
\ln\theta_1\approx\int_{1/|\overline{\lambda}|}^{\sqrt{|\overline{\lambda}|/2}}dr
\left(\frac{|\overline{\lambda}|}{2r^2}\right)^{1/3}\approx 3r^*.
\end{equation}

(iii) Using Eq.\ (\ref{eq3b.20}) with $\eta=0$ we find
\begin{equation}
\label{eq3b.35}
y\left(r\right)\sim\cot\left(r+\varphi\right)+
\frac{\overline{\lambda}}{4r^2}\sin^2\left(r+\varphi\right).
\end{equation}
The first term in this expression stems from the oscillations of the wave function,
while the second one describes the decrease ($\overline{\lambda}<0$) of
the wave function envelope from the center of the sample outwards. The second
contribution $\ln\theta_2$ is given by the principal value of the integral $\int ydr$:
\begin{equation}
\label{eq3b.40}
\ln\theta_2\approx -\mbox{\text P}
\int_{\sqrt{|\overline{\lambda}|/2}}^{L}y\left(r\right)dr\approx r^*/4.
\end{equation}

Adding the two contributions we establish the relation between
$\overline{\lambda}$ and $\theta$:
\begin{equation}
\label{eq3b.42}
\ln\theta=\frac{13}{4}r^*=\frac{13}{4}\sqrt{\frac{|\overline{\lambda}|}{2}},
\end{equation}
verifying the self-consistency of the assumption
$|\overline{\lambda}|\gg 1$ made at the beginning of this subsection.
 
The dimensionless integral in the saddle point action is also evaluated separately
in the regions (i)-(ii) and (iii):
\begin{mathletters}
\label{eq3b.43}
\begin{equation}
\label{eq3b.43a}
\int_0^{r^*}r^2v^2dr=3r^{*3}=\frac{192}{13^3}\ln^3\theta\,\,\,\,\,\,(\text{regions (i)
and(ii)})
\end{equation}
and
\begin{equation}
\label{eq3b.43b}
\int_{r^*}^{L}r^2v^2dr=r^{*3}/2=\frac{32}{13^3}\ln^3\theta\,\,\,\,\,\,(\text{region (iii)}).
\end{equation}
\end{mathletters}
Both contributions turn out to be proportional to the same power of $\ln\theta$
indicating that local resonances and Bragg reflection play equally important
roles in the formation of anomalously large wave function intensities.
Combining the results we obtain
\begin{equation}
\label{eq3b.44}
{\cal P}\left(\theta\right)\sim\exp\left\{-\frac{56}{2197}p_Fl\ln^3\theta\right\}.
\end{equation}

The envelopes of the wave function in the two regions are
\begin{mathletters}
\label{eq3b.45}
\begin{equation}
\label{eq3b.46}
\psi\left(r\right)\sim\frac{\psi\left(0\right)}{r}\exp\{-3r^{*2/3}r^{1/3}\}
\,\,\,\,\,(\text{for}\ \frac{1}{r^{*2}}<r<r^*)
\end{equation}
and 
\begin{equation}
\label{eq3b.47}
\psi\left(r\right)\sim\frac{\psi\left(0\right)}{r}\exp\{r^{*2}/4r-
\left(13/4\right)r^*\}\,\,\,\,\,(\text{for}\ r>r^*).
\end{equation}
\end{mathletters}
The normalization integral is given by
\begin{equation}
\label{eq3b.50}
N=\pi\psi^2\left(0\right)\left\{\frac{1}{3r^{*2}}+4Le^{-\frac{13}{2}r^*}
\right\}.
\end{equation}
When $L\gg \theta^2$ (with
more realistic boundary conditions this inequality will become
$L^3\gg\nolinebreak\theta^2$), the normalization integral is dominated by the
contribution from region (iii). Then
$\psi^2\left(0\right)\propto\nolinebreak\theta^2$, and
we finally obtain with exponential accuracy
\begin{equation}
\label{eq3b.60}
{\cal P}\left(t\right)\sim\exp\left\{-\frac{7}{2197}\left(p_Fl\right)\ln^3t
\right\}.
\end{equation}
Of course, the separation into regions $r<r^*$ and $r>r^*$ is approximate. It
is possible that the crossover region $r\sim r^*$ gives a contribution of the
same order of magnitude as the two asymptotic regions (ii) and (iii). Thus the
number $7/2197\approx 3.2\times 10^{-3}$ can only be considered as an
order-of-magnitude estimate of the coeficient $\kappa$ introduced in Section
\ref{sec2}. It must be mentioned that convergence to the asymptotic form of
Eq.\ (\ref{eq3b.60}) is extremely slow because of the stringency of the
requirement $r^*=\frac{4}{13}\ln\theta\gg 1$.

In the opposite regime $L\ll\theta^2$, $\psi^2\left(0\right)$ is close to its
maximal value $\sim 1$, and therefore its $\theta$ dependence is very weak. On
the other hand, similarly to the one-dimensional case, $\theta^2$ becomes
proportional to the electric response time $\tilde{t}$, leading to the 
distribution of $\tilde{t}$ having the form identical to Eq.\ (\ref{eq3b.60}),
\begin{equation}
\label{eq3b.70}
{\cal P}\left(\tilde{t}\right)\sim\exp\left\{-\kappa \left(p_Fl\right)
\ln^3\tilde{t}\Delta\right\}.
\end{equation}

Both Eq.\ (\ref{eq3b.60}) and Eq.\ (\ref{eq3b.70}) differ significantly from
the corresponding $\sigma$-model results \nolinebreak\cite{EfFal2,Kh3}. We
discuss the possible origins of this difference in the next section.

\section{Discussion}
\label{disc}

\subsection{General remarks.}

The main result of the work presented in this paper is that statistics of
seldom occurring events in disordered electronic systems can be successfully
studied using a variant of the optimal fluctuation method. All previous
approaches to the problem have been based on various formulations of the
nonlinear $\sigma$-model, and they invariably seem to require an extension of
the $\sigma$-model to, and sometimes beyond, its limits of validity. In
retrospect, this situation seems rather natural. Indeed, the success of the
$\sigma$-model in describing a wide variety of phenomena in chaotic and
disordered systems can be traced to the fact that most such phenomena are
quasiclassic in nature and therefore their most appropriate description is in
terms of an expansion into sums over quasiclassical trajectories. Calculation
schemes based on non-linear $\sigma$-models allow one to perform such
summations in a very efficient manner by introducing collective variables
($Q$-matrices) that change slowly in space \cite{Wegner,Efetov}.
When low-order moments and correlation functions are described the
integrals over $Q$-matrices get almost equal-weight contributions from the
whole degenerate (or near-degenerate) manifold \cite{Wegner,Efetov} of allowed
$Q$-matrix configurations. This effectively corresponds to probing a large
number of disorder configurations.

An investigation of the statistics of rare events, on the other hand, presents
quite a different type of problem. The configurations of the random potential
that give rise to such events come from a small subset of all possible
configurations. This circumstance was reflected in the approach developed in
Ref.\ \cite{Kh1} --  the large parameter $t$ served to strongly break the
degeneracy of the realizations of the $Q$-matrices, and only the vicinity of a
saddle point of the integral over $Q$'s was shown to give a significant
contribution to the probability of observing large $t$. What was effectively
encoded in such a scheme was the fact that the purpose of the calculation was
to select those special potentials that are capable of producing a given value
of $t$, and a corresponding selection of the relevant configurations of $Q$'s
was a computational tool allowing to do that.

The disadvantages of such an approach are obvious:

(i) First, the assumption that
the motion of electrons  can be described entirely in quasiclassical terms
imposes certain restrictions on the types of the potentials over which
averaging is performed. As follows from the calculations presented here, in three
spatial dimensions there exists a region $r\sim r^*$ where the solution of the
Schr\"{o}dinger equation cannot be obtained in the quasiclassical approximation.
As a result, the $\sigma$-model fails to recognize the existence of this region
and it also misses entirely the local resonance formed at $r<r^*$. 

(ii) Second, even
when the $\sigma$-model calculations succeed in correctly -- albeit
implicitly -- identifying the relevant disorder configurations  (as in one- and
two-dimensional samples) they do not always produce correct answers because
contributions from the short-wavelength degrees of freedom (massive
modes) eliminated in the
transformation from the fast to slow variables are missed.

\subsection{The direct optimal fluctuation method.}

It is not immediately evident, however, that a relatively na\"{\i}ve approach based
on the direct search for an optimal fluctuation should be more reliable. In
order for it to work, the probability of observing a large value of $t$ must be
determined by a sum over disorder configurations which all come from a single
compact region of the configurations space. The potentials forming this region
differ only slightly from some optimal configuration which corresponds to the
saddle point. Although we
have not proved in this work that the saddle point identified here gives the
dominant contribution to the functional integral, ``the preponderance of
evidence" based on the comparison of our results with those obtained using the
$\sigma$-model would indicate that this is indeed the case. 

The starting point of our qualitative analysis is the observation that, apart from the
difference in the cutoff scale, our variant of the optimal fluctuation method
(the ``direct" optimal fluctuation method) reproduces  identically in the
two-dimensional and one-dimensional cases the $\sigma$-model results for the
unitary ensemble. Assuming that more than a chance coincidence is involved, it
is reasonable to conclude that the optimal configurations of disorder found
in this work are the same as the ones that are responsible for the saddle point
of the $\sigma$-model. In other words, there must exist a one-to-one
correspondence between the saddle points of the theory defined by Eqs.\
(\ref{eq3.30}) and the saddle points of the $\sigma$-model. Assuming the
existence of such a correspondence we will try to elucidate the origins of the
$\sigma$-model results by analyzing the corresponding optimal
configurations of the potential $U\left({\bf r}\right)$.

Before proceeding with this analysis we would like briefly to address the
question of the stability of the saddle point described by Eqs.\
(\ref{eq3.40}). In the three-dimensional case the dominant contribution to the
saddle point action ${\cal A}_0$ comes from small distances where the optimal
potential is much larger than its typical Gaussian fluctuation, and stability
with respect to fluctuations does not pose a problem. In two dimensions,
however, the outlying regions of the optimal configuration -- where the
magnitude of the potential vanishes as $1/r$ -- must be taken into account. In
order to argue that fluctuations do not destroy the saddle point we
note that although we find it convenient to write the Gaussian distribution
function ${\cal W}\left[U\right]$ for the potential in the coordinate
representation, it can be written in any orthonormal basis $\left\{f_n\left({\bf
r}\right)\right\}$. A typical amplitude of a dimensionless basis function $f_n$ in a
typical configuration $U\left({\bf r}\right)$ is $1/\sqrt{\nu_2D}$. Let us now
choose one of the basis functions, say $f_0\left({\bf r}\right)$, to be proportional
to the optimal solution $v\left(r\right)$ given by Eq.\ (\ref{eq3c.10}). 
Using the normalization condition $\int d{\bf r} f_0^2\left({\bf r}\right)=1$ we obtain
\begin{mathletters}
\label{eq4.10}
\begin{equation}
\label{eq4.11}
f_0\left(r\right)\approx\sqrt{\frac{1}{\pi\ln\left(L/r_0\right)}}
\frac{\sin 2r}{r},
\end{equation}
and
\begin{equation}
\label{eq4.12}
v\left(r\right)\sim \frac{\ln\theta}{\sqrt{\ln\left(L/r_0\right)}}f_0
\left(r\right).
\end{equation}
\end{mathletters}
Thus the optimal fluctuation has a much larger amplitude than the typical one
as long as $\ln^2\theta\gg\displaystyle{\frac{\ln\left(L/r_0\right)}{\nu_2D}}$. This
condition, of course, is just a natural requirement for the validity of the
optimal fluctuation method $|\ln{\cal P}|\gg 1$.

It remains to be shown, however, that other components of a typical fluctuation
$U\left({\bf r}\right)$ (i.e. those orthogonal to $f_0$) do not destroy the
saddle point. A rigorous investigation of the fluctuations around the saddle
point defined by Eqs.\ (\ref{eq3.40}) will be the subject of a forthcoming
publication. Nevertheless, a plausible argument in favor of the stability of 
this saddle point can be made based on the following observation. Since the
appearance of anomalously localized states due to Bragg reflection is a
phenomenon analogous to the emergence of a band structure in a periodic lattice,
the suppression of this effect by fluctuations of the potential is equivalent
to the localization transition which destroys the band structure in ordinary
periodic lattices. Therefore $L<L_c$ seems to be a sufficient condition for the 
stability of the saddle-point solution in the two-dimensional case.

\subsection{The asymptotics of the distribution functions.}

\subsubsection{The one-dimensional case.}
\label{disc1d}

We will now try to use the physical intuition afforded by the optimal
fluctuation concept to compare the $\sigma$-model results of Eqs.\
(\ref{eq1.10}) and (\ref{eq1.30}) with the distribution functions Eqs.\
(\ref{eq3a.70}), (\ref{eq3c.50}), (\ref{eq3b.60}) and (\ref{eq3b.70})
derived in
the preceding Section. The first question that can be easily answered is why 
the distribution function for the electric response times in the
one-dimensional case is log-normal instead of a power law (i.e.,
$\exp\left(-C_1 \ln \tilde{t}\right)$) obtained by a na\"{\i}ve  extrapolation of
the $\exp\left(-C_d\ln^dt\right)$ dependence to $d=1$. In one and two
dimensions the optimal configurations found in the preceding Section are
``global", i.e. the integration in the saddle-point action $\int v^2d{\bf r}$ must be
extended over the whole sample. In contrast, in the three-dimensional case the
optimal fluctuation, even with the oscillating tail included, is local, so that the
above integral converges at large distances. It is well known \cite{Lifshitz} that
distribution function tails of the $e^{-\ln^d\tilde{t}}$ type usually appear as
probabilities of optimal fluctuations confined to a finite volume. Thus in
order to explain the ``anomaly" in the one-dimensional case we have to
understand why a local fluctuation of the random potential necessary to achieve
a given value of $\tilde{t}$ has a lower probability  than the "global" one 
proposed in Ref. \cite{Kh3} and rederived in Section \ref{sec3a}. A local
fluctuation of the potential leading to a large value of $\tilde{t}$ would have
to be able to support a narrow resonance. Assuming as in Section \ref{sec2} a
rectangular shape for the potential barrier responsible for the formation of
the resonance we can repeat the calculation presented there almost verbatim
except that $t$ must everywhere be replaced by $\tilde{t}\Delta$. We then 
obtain $\ln {\cal P}\approx\displaystyle{\frac{2}{3\sqrt{3}}\left(p_Fl\right)
\ln\tilde{t}\Delta}$. In order for this behavior to dominate we must have 
$p_Fl\ln\tilde{t}\Delta\ll\left(l/L\right)\ln^2\tilde{t}\Delta$, or
\begin{equation}
\label{eq4.30}
\ln\tilde{t}\Delta\gg p_FL.
\end{equation}
However, from Eq.\ (\ref{eq2.65}) we see that the corresponding values of $b$ --
which determine the size of this local fluctuation of the potential -- become
larger than the length of the sample $2L$. This is clearly unphysical. Thus
despite a slower $\tilde{t}$ dependence of the probability of local resonances,
the same value of $\tilde{t}$ has a much larger probability to be produced by
an accidentally formed Bragg mirror for all reasonable values of $\tilde{t}$.

Note also that $\ln\tilde{t}\Delta\sim p_FL$ corresponds to $v\sim 1$ (or
$U\sim E_F$), which invalidates the perturbative expansion of Section
\ref{sec3a}. Essentially, the values of $\ln\tilde{t}\Delta$ of the order of $p_FL$
or larger correspond to a trivial case: a sample is insulating because the
potential is larger than $E_F$ almost everywhere except for a small island in
the center where almost all the weight of an eigenstate at $E_F$ is
concentrated. In this regime the distinction between ``global" and local
fluctuations of the potential gets blurred.

It is interesting to note that a calculation based on the ballistic 
$\sigma$-model performed in Ref. \cite{Kh3} for {\em quasi}-one-dimensional
conductors indicates the existence of a crossover from log-normal to power-law
distribution at $\ln\tilde{t}\Delta\sim\displaystyle{\frac{L}{l}}$. Although
we have not investigated the quasi-one-dimensional case here, it is
likely that the physical picture of the interplay between the ``global" and
local fluctuations discussed above should not be much different. If that is
the case, then it is probable that the crossover found in Ref. \cite{Kh3} has
as its underlying cause the same mechanism of local fluctuations becoming
comparable in size to the length of the sample. This hypothesis, however, 
leaves unexplained the difference in scales at which the crossover occurrs --
$\displaystyle{\frac{L}{l}}$ in Ref. \cite{Kh3} as opposed to $p_FL$ in the
argument presented above. It is possible that the quasi-one-dimensional case
brings in some new features that are not recognized by the estimates based on
the purely one-dimensional model. It should be mentioned, however, that none
of the variants of the $\sigma$-model can provide an adequate description of
the effects associated with local resonances, and it is possible that a
$\sigma$-model estimation of the crossover scale may not be entirely reliable.

\subsubsection{Two-dimensional conductors.}

A rather comprehensive discussion of the results obtained by the direct optimal
fluctuation method in the two-dimensional case and their counterparts
established using the $\sigma$-model formalism has already been presented in
Section \ref{sec3b}. Here it seems appropriate to briefly reiterate the following 
two main points of that discussion. First, it comes as no surprise that
when massive modes are taken into account, the short distance cut-off scale in
the logarithm determining the system size dependence of $\ln{\cal P}$ becomes
of the order of the electron wavelength $p_F^{-1}$ rather than the mean free
path $l$. This leads, however, only to a small relative change in the
$\sigma$-model result. 

On the other hand, the apparent ensemble independence of the distribution
function asymptotes obtained by our method stands in a striking contradiction
with the letter and the spirit of the $\sigma$-model and is quite enigmatic.

\subsubsection{The three-dimensional case.}

The same as in the one-dimensional case issue of whether the $\sigma$-model 
calculations can correctly handle
the role of fluctuations of the potential with the amplitude larger than the Fermi
energy arises in connection with the $\ln^2 t$ asymptotics of the distribution
of wave function amplitudes obtained in Ref. \cite{EfFal2} for the {\em
three-dimensional} case. Derivation of the $\sigma$-model involves linearization
of the spectrum near the Fermi energy. As a result, there is no intrinsic
scale in the model that would relate the amplitude of the fluctuating potential 
to the electron energy. The only scale is provided by the dispersion of the
fluctuations of the potential $\displaystyle{\frac{1}{\pi\nu_d\tau}}$. This
limitation of the model does not affect the computation of probabilities of
typical events, or of the averages that are dominated by such events, because
typical potentials are small compared to the Fermi energy. However, a problem
arises when rare large-amplitude fluctuations of the potential $U\left({\bf r}
\right)$ become dominant. The $\sigma$-model cannot detect the existence of 
classical turning points around which the quasiclassical approximation breaks
down. Thus a possible explanation of the result of Falko and Efetov \cite{EfFal2}
is that within the $\sigma$-model approach the classical turning point at $r^*$ is
missed, and the Bragg mirror is effectively assumed to persist until distances of
the order of the mean free path $l$. In the notation of Section \ref{sec3c} that
would correspond to $|\overline{\lambda}|\sim l\ln\theta$, and then ${\cal A}_0
\propto\int_l^Lv^2r^2dr \propto\displaystyle{\frac{|\overline{\lambda}|^2}{l}}
\propto l\ln^2\theta$, leading to $\ln{\cal P}\sim -\left(p_Fl\right)^2\ln^2
t$ which coincides with the answer obtained by Falko and Efetov. The reason 
that such a cut-off scheme leads to a higher estimate for the
probability ${\cal P}$ is that it corresponds to an incorrect solution of the
Schr\"{o}dinger equation in the region of large potentials $U>E_F$ which
overestimates the rate of growth of the wave function amplitude towards the 
center of the sample. 

Another discrepancy between our results and those obtained with the help of
nonlinear $\sigma$-models in the three-dimensional case is the difference
in powers of $\left(p_Fl\right)$ in  the exponents in Eqs.\ (\ref{eq1.10}) and
(\ref{eq3b.70}). We believe that this discrepancy has the same origin as the
difference between $\displaystyle{\ln\frac{L}{l}}$ and $\displaystyle{\ln
\frac{L}{r_0}}$ in the two-dimensional case -- the error introduced by using the
mean free path $l$ to determine the cut-off scale. It is interesting to note
that while the cut-off procedure used in the diffusive $\sigma$-model approach of
Ref. \cite{Mirlin} is capable of producing only an order-of-magnitude estimate
$C_3\sim\left(p_Fl\right)^2$, the ballistic $\sigma$-model of Muzykantskii and
Khmelnitskii \cite{Kh2,Kh3}, while giving an illusion of computing the coefficient
$C_3$ exactly ($C_3=\displaystyle{\frac{\pi}{9\sqrt{3}}\left(p_Fl\right)^2}$),
nevertheless leads to the same extra power of $\left(p_Fl\right)$. To explain this
seemingly paradoxical situation we will first examine the cut-off procedure
employed in Ref. \cite{Mirlin}. It is based on the condition, pointed out in Ref.
\cite{Kh1}, that in order for the diffusive $\sigma$-model to be applicable, the
spatial gradients of the $Q$-matrix components cannot exceed $1/l$. Explicitly,
for the calculation performed in Ref. \cite{Mirlin} this condition reads
\begin{equation}
\label{eq4.40}
\left|\frac{d}{dr}\ln\lambda_1\right|<\frac{1}{l},
\end{equation}
where $\lambda_1$ parametrizes the non-compact bosonic sector of the
$Q$-matrix \cite{Efetov} (not to be confused with the Lagrange multiplier
$\lambda$ used throughout this work). The distance $l_*$ where this condition
is violated is used in Ref. \cite{Mirlin} as a short-distance cutoff.
It was conjectured in Ref. \cite{EfFal2} and later confirmed in Ref.
\cite{Mirlin2} that spatial structure of the saddle-point solution for 
$\lambda_1$ mimics the envelope of anomalously localized states described 
by the saddle point of the $\sigma$-model. Therefore the optimal configuration
of $\left(d/dr\right)\ln\lambda_1\left(r\right)$ corresponds to the
non-oscillating part of
$y\left(r\right)$ in the direct optimal fluctuation method.

This correspondence
allows us to see directly what effect an artificial short-distance cutoff at $l_*$
would have in our approach. From Eq.\
(\ref{eq3b.35}) we find $l_*^2\sim l|\overline{\lambda}|$. 
Introducing such a cut-off at $l_*$  into
Eq.\ (\ref{eq3b.40}) we obtain $|\overline{\lambda}|\sim l_*\ln\theta$ and
therefore
$l_*$ is estimated as $l_*\sim l\ln\theta$. If then the integral determining the
saddle-point action ${\cal A}_0$ is also cut off at $l_*$, we obtain 
\[
{\cal A}_0\sim l\int_{l_*}^Lv^2r^2dr\sim l\frac{|\overline{\lambda}|^2}{l_*}
\sim l^2\ln^3\theta,
\]
or $\left(p_Fl\right)^2\ln^3\theta$ in conventional units, leading to the incorrect
answer that was already quoted in the Introduction.

It is thus evident
that distances shorter than $l_*$ give an important contribution to the
saddle-point action, and this contribution cannot be accounted for by the 
diffusive nonlinear $\sigma$-model. It should be emphasized that excluding the
short-distance contribution in this way leads to a significantly smaller estimate
for the probability of observing a given (large) value of $t$. This can be
understood by using the following argument. The wave function
amplitude grows substantially between $r\sim l_*$ and $r\sim 1$. Neglecting
this growth leads to a need for a faster increase in $t$ between $L$ and $l_*$
which can only be achieved by means of a ``costly" boost in the amplitude of
the Bragg mirror. As a result, non-optimal configurations of the random potential
are selected.

Turning now to the calculation performed with the help of ballistic nonlinear 
$\sigma$-model for the three-dimen-sional case in Ref. \cite{Kh3} we notice that 
the distance $r_*$ which separates the ``reaction" and ``run-out" zones
is of the same order as $l_*$. The contribution to the saddle point action from the
``run-out" (diffusive) zone leads to the already quoted
$\displaystyle{\frac{\pi}{9\sqrt{3}}\left(p_Fl\right)^2}$ value for $C_3$,
while the ``reaction" zone produces a contribution that has one less power of the
large logarithm $\ln\theta$ and is thus neglected. In contrast, in the 
calculation presented in Section \ref{sec3c} of this paper, the contribution 
from distances of the order of $r^*\sim\ln\theta\ll l_*$ dominates the saddle 
point action and leads to a larger estimate (Eq.\ (\ref{res3D})) for the 
probability of observing anomalously high values of $t$ in the 3-dimensional 
case than the one obtained in Ref. \cite{Kh3}. Thus, the ballistic
generalization of the nonlinear $\sigma$-model is also not capable of detecting
the existence of the scale $r^*$ at which the quasiclassical approximation
breaks down. Moreover, since the calculation within the framework of the
ballistic $\sigma$-model does not involve any ultraviolet divergencies that
would necessitate a short-distance cut-off as in the case of the diffusive
$\sigma$-model, it appears plausible that the ballistic variant of the model
is equivalent to introducing an ultraviolet regularization into the theory.

\subsection{Prelocalized wave functions.}

To complete the comparison of the results obtained by the direct optimal
fluctuation method and those found using the nonlinear $\sigma$-models we
now turn to the issue of the shape of the envelope of anomalously localized
states that are responsible for the large-$t$ tails of the distribution functions
${\cal P}\left(t\right)$. In Ref. \cite{EfFal2} it was found that in two
dimensions such states are characterized by amplitudes decaying outwards 
according to a power law:
\begin{equation}
\label{eq4.60}
\left|\psi\left({\bf r}\right)\right|^2\sim \left(\frac{l}{r}\right)^{\ln
t/\ln\left(L/l\right)}.
\end{equation}
This result has to be compared with Eq.\ (\ref{eq3c.35}) which can be
rewritten as
\begin{equation}
\label{eq4.70}
\left|\psi\left({\bf r}\right)\right|^2\sim\frac{\psi^2\left(0\right)}{r}
\left(\frac{r_0}{r}\right)^{\ln t/\ln\frac{L}{r_0}}.
\end{equation}
Apart from the difference in the cut-off scale ($l$ vs. $r_0$) which was
discussed at length above, Eq.\ (\ref{eq4.70}) contains an extra $1/r$ factor
in the denominator. It is simply a consequence of the idealized model adopted
here with its circularly symmetric boundary conditions. In a more realistic
model the optimal wave function would become a superposition of different
angular momentum eigenstates. The $1/\sqrt{r}$ behavior of the circularly
symmetric component would be cancelled in such a superposition. Taking into
account fluctuations around the saddle point would also have the effect of
suppressing the $\left(1/\sqrt{r}\right)^2$ factor in the spatial dependence
of the {\em averaged} envelope of the optimal solution. Note also that Eq.\
(\ref{eq4.60}) does indeed describe the averaged envelope of anomalously
localized states.

In three-dimensional samples the states with anomalously high local amplitudes
were found in Ref. \cite{EfFal2} to have the following envelope:
\begin{equation}
\label{eq4.80}
\left|\psi\left({\bf r}\right)\right|^2\sim
\exp\left\{-A\left(1-\frac{l}{r}\right)\right\},
\end{equation}
where $A$ is a constant which in the leading logarithmic approximation is
equal to $\ln t$. Comparing this to Eq.\ (\ref{eq3b.45}) we see, in accordance
with the discussion above, that the $\sigma$-model result gives the correct
functional form $e^{\text{Const.}/r}$ only for the region of large $r$ which
corresponds to the region of space where the optimal potential forms a Bragg
mirror. The estimate $l\ln t$ for the constant in the exponential, however, 
contains an extra factor of $l$ compared to Eq.\ (\ref{eq3b.47}) which is
a consequence of the choice of the cut-off scale made in Ref.
\nolinebreak\cite{EfFal2}.
The region of the stretched exponential decay of the wave function envelope
described by Eq.\ (\ref{eq3b.46}) is missed in the $\sigma$-model calculation
entirely. As in the two-dimensional case, the extra powers of $r$ in the
denominators of Eqs.\ (\ref{eq3b.45}) are a consequence of the artificial
rotational symmetry. 

It was conjectured in Ref. \cite{EfFal2} that these high-amplitude states have
a complicated ``snake-like" structure at short distances. The conjecture was
based on the fact that the method employed in Ref. \cite{EfFal2} was applicable
to amplitudes as high as
\begin{equation}
\label{eq4.90}
t\lesssim\displaystyle{\frac{Vp_F^{d-1}}{l}},
\end{equation}
rather than a na\"{\i}ve $t\lesssim\displaystyle{\frac{V}{l^d}}$ expected from a
cut-off at $l$. We have found no evidence for such behavior here.
Since the solutions of the saddle point equations were assumed to be
rotationally invariant from the outset, this cannot be regarded as a
conclusive evidence against such a scenario. However, potentially more
important is the fact that we did not encounter any limitations on the
possible values of $t$ analogous to Eq.\ (\ref{eq4.90}). Analysis of the
fluctuations around the saddle point defined by Eqs.\ (\ref{eq3.40}) as well
as an investigation of the possibility for non-rotationally-invariant
solutions of these equations is needed to settle the issue conclusively.

\subsection{Are the $\sigma$-model results universal?}

Finally, we would like to make a few remarks concerning the issue of
universality.
An important consequence of the dominant role played in three dimensions by
large local fluctuations of the potential that are responsible for the
formation of local resonances is the non-universal character of the distribution
functions derived in Section \ref{sec3c}. Indeed, such large-amplitude
configurations of the potential can only be optimal if they are not too
``expensive" -- i.e. if their action ${\cal
A}$ is not too high -- compared with the low-amplitude ``global"
fluctuations of the potential (Bragg mirrors). Generalizing the distribution
of the random potentials to
\begin{equation}
\label{eq4.100}
{\cal W}_n\left[U\right]\propto\exp\left(-\frac{1}{2\sigma}\int 
U^{2n}\left({\bf r}\right)d^d{\bf r}\right),
\end{equation}
where $\sigma$ is the dispersion and $n\ge 1$, we find through a model
calculation similar to the one performed in Section \ref{sec2} that forming states
with large amplitudes at some ${\bf r}$ by means of local resonances always
leads to $\ln{\cal P}\propto -\ln^dt$. For $n<d/2$ the ``cost" of a corresponding
Bragg mirror can be estimated from an appropriate generalization of
Eqs.\ (\ref{eq3.50}), and it also leads to a
$\ln^dt$ dependence, as was demonstrated by a detailed study of Gaussian
($n=1$) distribution in three dimensions. Thus local large-amplitude fluctuations
of the potential are always important for small enough ($<d/2$) values of $n$.
In the marginal case $n=d/2$ the weight of the Bragg mirror is given by
\begin{equation} 
\label{eq4.110}
\ln {\cal P}\propto
-\frac{\ln^dt}{\left(\displaystyle{\ln\frac{L}{r_0}}\right)^{d-1}},
\end{equation}
making it a much
more probable way of achieving a large local wave function intensity $t$. When
$n>d/2$ the probability of the corresponding Bragg mirror has a faster than
$\ln^dt$ dependence on $\ln t$, but it is compensated by a power of the system
size in the denominator:
\begin{equation}
\label{eq4.120}
\ln{\cal P}\propto -\frac{\ln^{2n}t}{L^{2n-d}}.
\end{equation}
This situation is realized in the Gaussian case when $d=1$, and it was
already discussed in detail. Therefore the seemingly universal character of the
log-normal behavior of the tails of the distribution functions is tied to the
assumption -- which is a starting point in the derivation of the $\sigma$-model
-- that the random potential has a Gaussian distribution.

\section{Conclusions and open questions.}
\label{sec5}

The most important problem arising from the results of the present study is
the need to explain the fact that the asymptotics of the distribution
functions derived using the direct optimal fluctuation method do not exhibit any
dependence on weak magnetic fields. This feature of our answers must be
contrasted with {\em all} the previous calculations performed
in the framework of nonlinear $\sigma$-models in which it is quite obvious that
when Cooper modes acquire a mass as a consequence of broken invariance
with respect to time reversal, the number of independent components of the
$Q$-matrices changes and that has a profound impact on the results. There are
two alternative resolutions of this contradiction. 

One possibility is that our assumption that rotationally invariant solutions of the
saddle-point equations dominate the saddle-point action is not valid in the
absence of magnetic field. A more detailed study of the properties of Eqs.\
(\ref{eq3.40}) will be the subject of a future publication.

The second possible
explanation is that there is some internal inconsistency in the $\sigma$-model
which causes it in certain circumstances to overcount the number of degrees of
freedom. In our view, this possibility must be taken seriously.

Whether or
not the fluctuations around the saddle point can change the leading order terms
in $\ln t$ is also one of the questions that were left beyond the scope of this
work. A peculiar disagreement in the crossover scale from $\ln^2t$ to $\ln
t$ asymptotics of $\ln{\cal P}$ between the purely one-dimensional and
quasi-one-dimensional cases which was noted in Section \ref{disc1d} makes it
rather desirable to extend our method to the quasi-1D and quasi-2D geometries.

To conclude, the main results of this work can be summarized as follows. We have
demonstrated that the optimal fluctuation method is a useful tool for the
investigation of the statistical properties of anomalous electronic eigenstates in
disordered two- and three-dimensional conductors. It was shown that in three
dimensions this method is preferable to the nonlinear $\sigma$-model because
the latter does not include effects associated with local resonances that can be
formed by the random potential. In the two-dimensional case the results obtained
by the optimal fluctuation method essentially coincide with the $\sigma$-model
results for the unitary ensemble of random potentials which we interprete as an
indication that the saddle point of the reduced nonlinear $\sigma$-model found
in Refs. \cite{Kh1,EfFal1,EfFal2,Mirlin,Kh3} corresponds to the same saddle point
of the full theory as the one that describes the optimal fluctuation of the
potential.

\section*{Acknowledgements}
One of the authors (I. S.) would like to thank N.~Win-green, H.~Li, D.~Fisher,
B.~Halperin and A.~Andreev for useful discussions. I.~S. also acknowledges
financial support from NSF grant DMR 9106237.

\appendix

\section*{}
\label{a1}

Introducing an auxiliary field $\chi$ as in the main text, we can represent 
${\cal N}_{\psi}$ as
\begin{equation}
\label{eqap1}
{\cal N}_{\psi}^{-1}=\int {\cal D}\psi {\cal D}\left(\frac{\chi}{2\pi}\right) 
e^{i\int d{\bf r}\chi\left({\bf r}\right)\left(\frac{\hat{\bf p}^2}{2m}+
U\left({\bf r}\right) -E\right)
\psi\left({\bf r}\right)}.
\end{equation}
Performing a $\pi/4$ rotation in the $\left(\chi,\psi\right)$ space 
\begin{equation}
\chi=\left(\psi_1+\psi_2\right)/\sqrt{2},\,\,\,\,\,\,\, \psi=\left(\psi_1-
\psi_2\right)/\sqrt{2},
\end{equation}
and introducing infinitesimal convergence factors we obtain
\begin{eqnarray}
\label{eqap2}
{\cal N}_{\psi}^{-1} & = & \int {\cal D}\left(\frac{\psi_1}{\sqrt{2\pi}}\right)
{\cal D}\left(\frac{\psi_2}{\sqrt{2\pi}}\right) \nonumber \\
& \times & e^{\frac{1}{2}i\int d{\bf r}
\{\psi_1\hat{O}^{\left(+\right)}\psi_1 - \psi_2\hat{O}^{\left(-\right)}
\psi_2\}},
\end{eqnarray}
where $\hat{O}^{\left(\pm\right)}=\frac{\hat{\bf p}^2}{2m} + 
U\left({\bf r}\right) - E \pm i0$.
Thus, up to a constant, ${\cal N}_{\psi}$ is equal to a symmetrized spectral
determinant:
\begin{eqnarray}
\label{eqap3}
{\cal N}_{\psi} & \propto & \sqrt{\det\left\{E-\hat{H}+i0\right\}\det\left\{E-
\hat{H}-i0\right\}} \nonumber \\
& = &
e^{\Re\text{Tr}\ln\left(E-\hat{H}+i0\right)}.
\end{eqnarray}
The variation of the logarithm in the exponent with respect to $U\left({\bf r}
\right)$ is equal to the real part of the Green's function at coinciding
arguments $\Re G\left({\bf r},{\bf r};U\right)$. This quantity probes the
whole band and is not sensitive to small changes in $U$. It therefore
can be treated as a constant (which we denote as $G$), so that 
${\cal N}_{\psi}$ becomes
\begin{equation}
\label{eqap4}
{\cal N}_{\psi}\propto e^{G\int d{\bf r}U\left({\bf r}\right)}.
\end{equation}
The background potential $\int d{\bf r}\,U\left({\bf r}\right)$ can be
absorbed into a redefinition of energies, which justifies the approximation
made in deriving Eqs.\ (\ref{eq3.40}).

It should also be noted that in the sigma model formalism the real part of the
Green's function $G\left({\bf r},{\bf r}\right)$ is effectively set to zero
under the assumption of an infinite symmetric band. Therefore any corrections
to the distribution function that may arise due to ${\cal N}_{\psi}\neq 1$ are
at any rate beyond the scope of the $\sigma$-model.

\begin{references}
\bibitem{redbook}{\em Mesoscopic Phenomena in Solids}, B. L. Altshuler, P. A.
Lee and R. A. Webb, eds., North Holland 1991.
\bibitem{AKL}B. L. Altshuler, V. E. Kravtsov and I. V. Lerner, in \cite{redbook},
and references therein.
\bibitem{Wegner}F. Wegner, Z. Phys. B {\bf 35}, 207 (1979).
\bibitem{Kh1}B. A. Muzykantskii and D. E. Khmelnitskii, Phys. Rev. B {\bf 51}, 5480
(1995).
\bibitem{Efetov}K. B. Efetov, Adv. Phys. {\bf 32}, 53 (1983).
\bibitem{EfFal1}V. I. Falko and K. B. Efetov, Europhys. Lett. {\bf 32}, 627
(1995).
\bibitem{EfFal2}V. I. Falko and K. B. Efetov, Phys. Rev. {\bf B52},17413 (1995).
\bibitem{Mirlin}A. D. Mirlin, Phys. Rev. {\bf B53}, 1186 (1996).
\bibitem{Kh3}B. A. Muzykantskii and D. E. Khmelnitskii, unpublished (preprint
cond-mat/9601045).
\bibitem{Kh2}B. A. Muzykantskii and D. E. Khmelnitskii, JETP Letters {\bf 62}, 76
(1995), [Pis'ma Zh. Eksp. Teor. Fiz. vol. 62, p. 68 (1995)].
\bibitem{AltPrig}B. Altshuler and V. Prigodin, JETP Letters {\bf 47}, 43 (1988),
[Pis'ma Zh. Eksp. Teor. Fiz. vol. 47, p.36 (1988)].
\bibitem{HalpLax}B. I. Halperin and Melvin Lax, Phys. Rev. {\bf 148}, 722 (1966).
\bibitem{ZittartzLang}J. Zittartz and J. S. Langer, Phys. Rev. {\bf 148}, 741 (1966).
\bibitem{note1}Such a formulation of the problem implies perfect rotational
symmetry and sidesteps the issue of the role of irregular boundary conditions.
However, as will become evident from the answers, the asymptotic form of the 
three-dimensional distribution is not sensitive to the boundary conditions at all.
The two-dimensional case is trickier and may require a separate study of the
stability of the solution with respect to the change in boundary conditions.
Nevertheless, since the $\sigma$-model calculations were performed for similar
geometries, the model adopted here is completely adequate for the purposes of
comparing different approaches.
\bibitem{note3}In the formalism of Falko and Efetov \cite{EfFal1}
the extra delta-function in the definition of ${\cal P}$ was 
essential in enabling the mapping of the problem onto the nonlinear 
$\sigma$-model. However, the saddle-point solution of the reduced variant of 
the model which was derived in Ref. \nolinebreak\cite{EfFal1} to describe
deviations from the universal Porter-Thomas statistics \cite{Haake} was shown
to involve only the variables parametrizing the bosonic sector of the theory. This
can be viewed as an indication that effects associated with the statistics of energy
levels are not important in the approximation corresponding to the saddle-point
treatment of the nonlinear $\sigma$-model. In the context of the direct
optimal fluctuation method this observation receives quite a natural
explanation. A given value of energy $E$ can be fine-tuned to become an
eigenvalue by a shift in the optimal configuration of $U\left({\bf r}\right)$
of the order of the mean level spacing $\Delta = \displaystyle{\frac{1}{\nu_dV}}$.
Such a shift can result in corrections to $\ln{\cal P}$ that are at most
$\sim E_F\tau$ -- much less then the leading contribution when $\ln t$ is
large.
\bibitem{Haake}Haake, {\em Quantum signatures of Chaos}, Springer-Verlag,
Berlin, New York 1991.
\bibitem{note2}In several cases the probabilities obtained by our method are {\em
smaller} than the correponding $\sigma$-model results. The relation between
the two sets of answers and the likely reasons for the differences between them
are discussed in Sections \ref{sec3} and \nolinebreak\ref{disc}.
\bibitem{Kamke}E. Kamke, {\em Differential Equations}, Chelsea Publ., NY 1971.
\bibitem{NumRec}W. H. Press {\em et.al}, {\em Numerical Recipes in C},
Cambridge University Press, Cambridge \nolinebreak 1988.
\bibitem{Mirlin2}A. D. Mirlin, unpublished (preprint cond-mat/9512095).
\bibitem{Lifshitz}I. M. Lifshitz, Soviet Phys. -- JETP {\bf 17}, 1159 (1963).

\end{references}
\end{document}